%% file: manuscript.tex
\PassOptionsToPackage{dvipsnames,table}{xcolor}

\documentclass[preprint,12pt]{elsarticle}

\usepackage{graphicx}
\usepackage{graphics}
\usepackage{flafter}
\usepackage[section,above]{placeins}

\usepackage{amssymb}
\usepackage{amsmath}
\usepackage{bm}
\usepackage{upgreek}
\usepackage{amsthm}
\usepackage{mathrsfs}
\usepackage{rotating}
\usepackage{tabularx}
\usepackage{amsfonts}
\usepackage{CJKutf8}
\usepackage{nicefrac}

\newcommand{\dd}{\,\mathrm{d}}

\newcommand{\PCTF}{\operatorname{PCTF}}
\newif\ifshowintrohints
\showintrohintsfalse
\usepackage[colorlinks=true,
            linkcolor=blue,
            citecolor=blue,
            urlcolor=blue]{hyperref}

\usepackage[capitalize,nameinlink,noabbrev]{cleveref}
\usepackage{todonotes}

\journal{Ultramicroscopy}

\begin{document}

\begin{frontmatter}



\title{Parallax Depth Sectioning and 3D Reconstruction in 4D-STEM} 

\author[aff:cornell]{Desheng Ma\corref{cor}}
\ead{dm852@cornell.edu}
\author[aff:cornell]{Chia-Hao Lee}
\author[aff:chem]{Zixiao Shi}
\author[aff:cornell,aff:kavli]{David A. Muller}
\author[aff:cornell]{Steven E. Zeltmann\corref{cor}}
\ead{steven.zeltmann@cornell.edu}
\cortext[cor]{Corresponding authors}

\affiliation[aff:cornell]{organization={School of Applied and Engineering Physics, Cornell University},
            city={Ithaca},
            postcode={14853},
            state={NY},
            country={USA}}
\affiliation[aff:chem]{organization={Department of Chemistry and Chemical Biology, Cornell University},
            city={Ithaca},
            postcode={14853},
            state={NY},
            country={USA}}
\affiliation[aff:kavli]{organization={Kavli Institute at Cornell for Nanoscale Science},
            city={Ithaca},
            postcode={14853},
            state={NY},
            country={USA}}

\begin{abstract}
\sloppy
Three-dimensional (3D) information is encoded in four-dimensional scanning
transmission electron microscopy (4D-STEM) through parallax between the virtual images formed at each detector pixel. 
In this paper we connect the encoding and retrieval of depth-dependent information in 4D-STEM to two other widely-used 3D imaging methods, tomography and light-field photography, and derive its 3D phase contrast transfer function.
Viewing the 4D data as a collection of angular slices yields a collection of sinograms, providing an intuitive visual representation of the 3D information transfer. 
Within the small-angle approximation, tilt-corrected bright-field (tcBF) depth sectioning is equivalent to non-iterative tomographic reconstruction.
Alternatively, the 4D data also can be mapped to light-field (or plenoptic) imaging, which acquires spatially resolved diffraction data in parallel using an array of lenslets.
Adapting a light-field imaging algorithm, we show that tcBF depth sections can be obtained by extracting two-dimensional slices from a single four-dimensional fast Fourier transform (4D FFT) of the dataset, yielding a substantial speedup when reconstructing a large number of depth slices.
These differing perspectives distinguish the geometric aspects of depth resolution from wave-optical effects that are present in phase-contrast imaging. 
Depth-resolved aberration-corrected bright-field (acBF) imaging yields improved 3D reconstructions compared to tcBF, removing contrast oscillations and reducing the variation in response with depth. 
We demonstrate acBF depth sectioning using simulated and experimental data and resolve distinct layers in stacked oxide films and nanoparticle assemblies.
These non-iterative volumetric reconstructions from a single 4D-STEM dataset show promise for improved imaging of thick, weakly-scattering samples and for reconstructing low-dose acquisitions.
\end{abstract}




\end{frontmatter}


\section{Introduction}\label{Introduction}
Conventional scanning transmission electron microscopy (STEM) forms an image by integrating selected parts of the diffraction plane during acquisition, as in bright-field (BF) imaging, annular bright-field (ABF) imaging, and high-angle annular dark-field (HAADF) imaging \citep{kirkland_2020_AdvancedComputingElectron}.
A segmented detector can be used to produce differential phase contrast (DPC) \citep{dekkers_1974_DIFFERENTIALPHASECONTRAST, rose_1977_Nonstandardimagingmethods}.
These imaging modes provide complementary contrast mechanisms for probing atomic structure, composition, and local electric and magnetic fields.

Four-dimensional STEM (4D-STEM) records the full convergent beam electron diffraction pattern (CBED) at every probe position using a pixelated detector, preserving the full momentum distribution of the scattered wave \citep{ophus_2019_FourDimensionalScanningTransmission}, leading to a phase-space  description of 4D-STEM \citep{rodenburg_1992_TheorySuperResolutionElectron}.
The 4D data can then be read in two ways: a diffraction pattern at each scan position, or a virtual image assembled from one detector pixel at every scan position. 
The differences between these virtual images contain information that an integrating detector discards but that can be retrieved computationally and used for new imaging and analysis methods \citep{ophus_2019_FourDimensionalScanningTransmission,ma_2026_Information4DSTEMWhere,varnavides_2026_RelaxingDirectPtychography}.

The simplest of these approaches is virtual imaging under any conventional integrating detector geometry, reproducing conventional STEM imaging modes after acquisition \citep{caswell-high-speed-2009,tate_2016_HighDynamicRange,ophus_2019_FourDimensionalScanningTransmission}.
Phase retrieval methods, such as ptychography, can recover object phase from interference between the transmitted and scattered waves. 
Under the weak phase object assumption (WPOA), a linear phase contrast transfer function (PCTF) exists and specifies which object spatial frequencies are recorded and transferred at each detector angle
\citep{rose_1977_Nonstandardimagingmethods,ma_2026_Information4DSTEMWhere}. 
With a defocused probe, the virtual images are predicted to be laterally shifted relative to one another. Reciprocity relates this shift to that of a TEM image under tilted illumination \citep{cowley_1969_ImageContrastTransmission}, and the amount of shift is determined by the aberration function of the probe \citep{lupini_2016_Rapidaberrationmeasurement,ma_2026_UsingAberrationsImprove}.
TcBF is a simple reconstruction approach for recovering a coherent phase contrast image by registering and summing the shifted virtual images, and acBF improves the contrast transfer by also correcting the full PCTF before summation \citep{ma_2026_Information4DSTEMWhere, ma_2026_UsingAberrationsImprove, ma_2026_WhyPtychographyWorksBetter}.
TcBF has also enabled low-dose imaging of specimens more than 500~nm thick
\citep{yu_2025_DoseEfficientCryoElectronMicroscopy}, and parallax has been used
to relax scan-sampling requirements in direct ptychography
\citep{yu_2025_DoseEfficientCryoElectronMicroscopy,varnavides_2026_RelaxingDirectPtychography,ma_2026_Information4DSTEMWhere}.

The different shifts reveal the parallax effect and provide insights into the 3D information and its recovery from a single 4D data set. 
Registering the images for one axial plane leaves features at other depths shifted and blurred. 
Therefore, by sweeping through the plane of reconstruction, a depth stack can be reconstructed from a single 4D-STEM acquisition \citep{zeltmann_2026_ImprovingLowDoseThreeDimensionalImaging}.
Such volumetric reconstructions can be used directly for locating features in 3D, or alternatively a reconstructed depth stack could be combined using a focus-selection rule, or a filter spanning several refocusing planes could retain a chosen depth range, as in optical volumetric-focus filtering \citep{dansereau_2015_LinearVolumetricFocus}.

Depth sectioning and its limits have been studied in through-focal STEM, extended-depth-of-field imaging, and scanning-confocal electron microscopy \citep{hovden_2011_ExtendedDepthField,behan_2009_Threedimensionalimagingoptical, xin_2010_ThreeDimensionalImagingAberrationCorrected}.
An early example of parallax refocusing in STEM was demonstrated by Seifer et al \citep{seifer_2021_FlexibleSTEMSimultaneous} and later extended in Refs. \citep{seifer_2026_Shadowmontageconebeam, kirchwegerAzimuthalSegmentImaging2025}.
Optical sectioning and post-reconstruction refocusing using Wigner distribution deconvolution ptychography was demonstrated by Yang et al \citep{yang_2016_SimultaneousAtomicResolution}, who additionally derived the shape of the 3D contrast transfer envelope.
Further work on the 3D CTF of ptychography was shown by Terzoudis-Lunsden et al \citep{terzoudis-lumsden_2023_ResolutionVirtualDepth} via analogy to a hypothetical phase-sensitive confocal instrument.

A sinogram, commonly used in tomography, provides an intuitive visualization of the 3D information. 
The position--angle sinogram places a projected profile from each angular view alongside the others, so the motion of an object feature traces a sinusoidal trajectory. 
In tomography, these profiles are acquired by changing the viewing angle and the original 3D object can be recovered from the sinogram. 
In 4D-STEM, the same information is acquired by scanning probe position and recording all detector angles at each position without rotating the specimen. 
This organization also connects 4D-STEM to a light field, a concept and implementation in light optics and photography for post-acquisition refocusing, which records both position and propagation direction instead of integrating over one particular direction at acquisition \citep{ng_2005_LightFieldPhotography,ng_2006_Digitallightfield}.
Light-field refocusing provides a computationally convenient method for depth sectioning. 
By viewing the full 4D data in Fourier space, the Fourier-slice theorem turns the same shift-and-sum refocusing used by tcBF into sampling a two-dimensional plane in the four-dimensional Fourier transform of the data \citep{ng_2005_Fourierslicephotography}.
Computing this transform once allows many depth sections to be reconstructed
without repeating the registration and summation. 

In this paper, we explore one-shot tcBF and acBF depth sectioning from a single 4D-STEM data set.
We first show how depth appears as parallax in a position--angle plot, or sinogram, and how registration and phase correction restore the coherent contrast at each depth level.
We use single- and two-atom simulations to separate the effects of image shift and phase transfer and show how tcBF and acBF differ in their manipulation of the position--angle space. 
We then show how a 4D-STEM acquisition can be connected to a lightfield camera despite their different physical configurations, and adapt a common reconstruction algorithm from the lightfield literature to accelerate the computation of tcBF depth stacks.
We finally derive the 3D PCTFs of tcBF and acBF depth sectioning to explain the axial response and its dependence on convergence angle and defocus. 
Reconstructions of a twisted SrTiO$_3$ bilayer and overlapping Pt nanoparticle grains demonstrate the method experimentally.

\section{Parallax Refocusing in 4D-STEM}
\label{sec:parallax_refocusing}

\subsection{Virtual images and shadow images}
\label{sec:virtual_images_sinograms}

\begin{figure}[!htbp]
    \centering
    \includegraphics[width=1.0\textwidth]{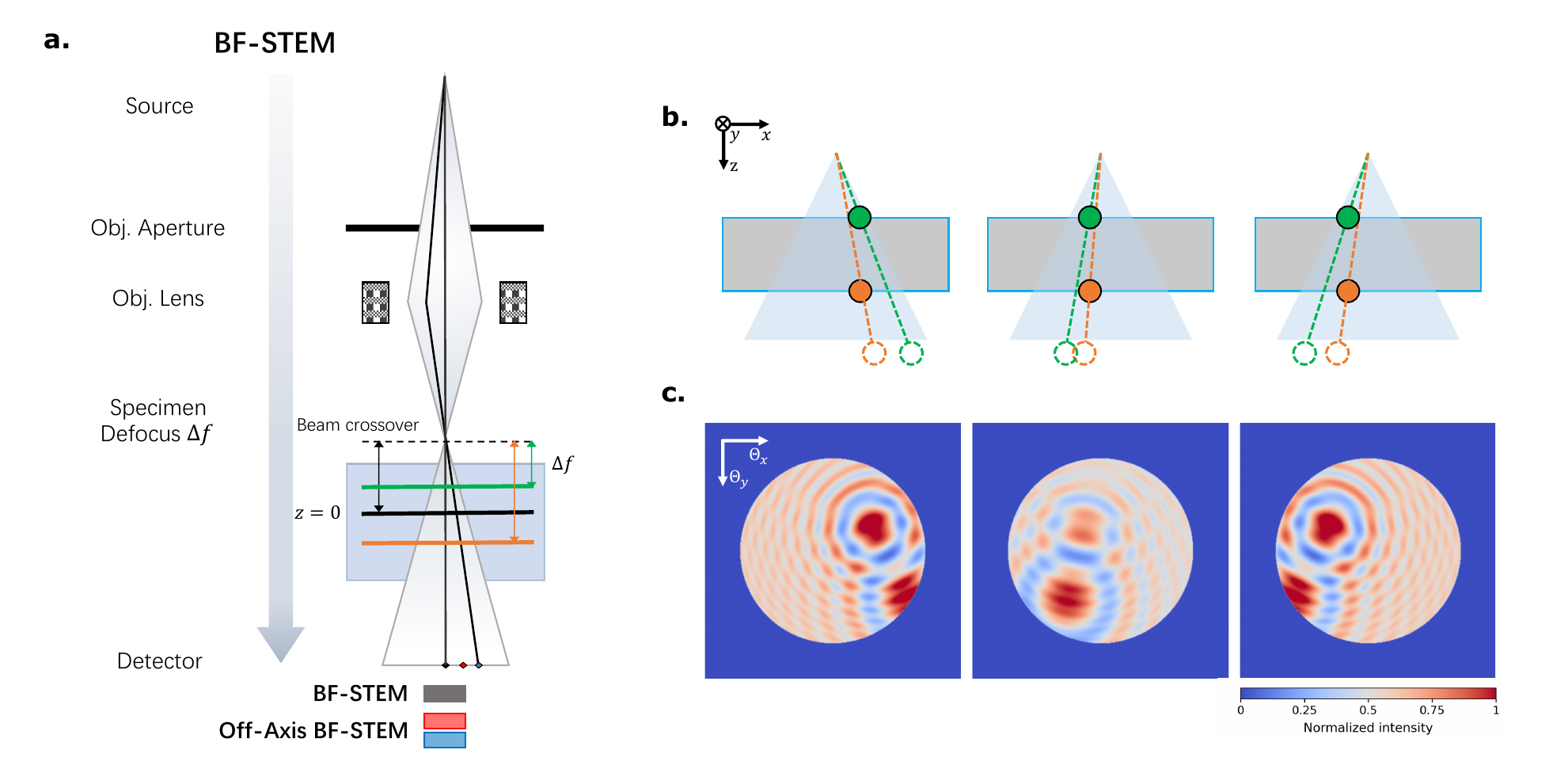}
    \caption{Parallax in defocused 4D-STEM. (a) In the schematic ray diagram, 
    a specimen away from the beam crossover is viewed by both on-axis 
    and off-axis BF rays. By reciprocity, off-axis virtual brightfield images are equivalent to conventional images with a corresponding beam tilt. These tilted rays intersect features at different positions depending on their depth in the sample. (b) The parallax effect. Features at different distance to the beam crossover are swept by the diverging probe at different rates as the beam scans. 
    (c) The corresponding diffraction patterns retain this depth-dependent shift. These effects encode 3D information into a 4D-STEM measurement.}
    \label{fig:parallax_schematic}
\end{figure}

Let $I(\mathbf r,\boldsymbol\Theta)$ be the measured intensity at probe position $\mathbf r=(x,y)$ and detector angle $\boldsymbol\Theta=(\Theta_x,\Theta_y)$.
Holding $\boldsymbol\Theta$ fixed and reading that detector pixel throughout the scan gives a virtual image. 
By reciprocity, the image from each detector pixel corresponds to an image from an equivalently tilted beam in conventional TEM \citep{cowley_1969_ImageContrastTransmission,ma_2026_Information4DSTEMWhere}.
For a defocused probe, an object feature moves between these images by an amount that depends on its depth via the parallax effect (Fig.~\ref{fig:parallax_schematic}).

When analyzing the contrast in 4D-STEM, we first subtract the object-independent background intensity
$I_0(\boldsymbol\Theta)$:
\begin{equation}
D(\mathbf r,\boldsymbol\Theta)
=I(\mathbf r,\boldsymbol\Theta)-I_0(\boldsymbol\Theta).
\label{eq:detector_resolved_contrast}
\end{equation}
This subtraction removes only the zero spatial frequency of a virtual image, which is never transferred by phase contrast imaging.
We use $\boldsymbol q=(q_x,q_y)$ for the lateral spatial frequency, conjugate to $\mathbf r$ under the kernel $\exp[-i2\pi\boldsymbol q\cdot\mathbf r]$.
The electron wavenumber is $k_0=2\pi/\lambda$ and the scattering vector is $\boldsymbol{\omega} = \lambda\boldsymbol q$ in Rose's notation \citep{rose_1977_Nonstandardimagingmethods}.
The axial spatial frequency $q_z$ is conjugate to $z$ and has the same unit of \AA$^{-1}$.
For a projected weak phase object, we can define a linear phase contrast transfer function (PCTF): 
\begin{equation}
\tilde D(\boldsymbol q,\boldsymbol\Theta)
=\PCTF(\boldsymbol q,\boldsymbol\Theta)
\tilde\phi(\boldsymbol q).
\label{eq:virtual_image_pctf}
\end{equation}
Here the tilde denotes Fourier transformation over probe position, and $\tilde\phi$ is the projected phase.
We first use this projected phase to define refocusing at a different axial planes.
Section~\ref{sec:3d_ctf} retains object depth explicitly to obtain the 3D PCTF response.

\subsection{Depth-dependent registration: tcBF}
Let the $z$ coordinate be the axial distance from the focal plane of the lens, with the beam propagating in the $+z$ direction (Fig \ref{fig:parallax_schematic}).
The defocus is referenced to the plane $z=0$, and defined so that positive $\Delta f$ places the beam crossover after the sample (i.e. at $z>0$) and $\Delta f = - C_{10}$. 
Thus, the signed effective defocus at a given plane is $\Delta f_{\mathrm{eff}}=\Delta f-z$, and the depth relative to the beam crossover is $\Delta z=z-\Delta f=-\Delta f_{\mathrm{eff}}$.
With the Fourier convention above and grouping the Ewald sphere term into the aberration phase $\chi=-k_0(\Delta f-z)|\boldsymbol\Theta|^2/2$, the Fourier shift factor in a virtual image is $\exp[i(2\pi\boldsymbol q)\cdot\boldsymbol\Theta(\Delta f-z)]$.
An image containing this factor has its features shifted by $-(\Delta f-z)\boldsymbol\Theta$.
Registration cancels this shift, and tcBF sums the registered images:
\begin{equation}
J_z^{\mathrm{tcBF}}(\mathbf r)
=\int_{|\boldsymbol\Theta|<\alpha}
D\!\left(\mathbf r-(\Delta f-z)\boldsymbol\Theta,
\boldsymbol\Theta\right)\dd^2\boldsymbol\Theta.
\label{eq:realspace_projection}
\end{equation}
Features at the reconstruction plane coincide among the shifted images, whereas features at other depths remain shifted and so are scrambled in the final image. 
Repeating this operation for different values of $z$ gives a three-dimensional reconstruction from a single acquisition.

The same operation can be expressed in Fourier space by Fourier transforming each virtual image, multiplying by a phase factor given by the Fourier shift theorem, summing over detector pixels, and inverse transforming:
\begin{equation}
\tilde J_z^{\mathrm{tcBF}}(\boldsymbol q)
=\int_{|\boldsymbol\Theta|<\alpha}
\tilde D(\boldsymbol q,\boldsymbol\Theta)
e^{-i(2\pi\boldsymbol q)\cdot\boldsymbol\Theta(\Delta f-z)}
\dd^2\boldsymbol\Theta.
\label{eq:tcbf_fourier_registration}
\end{equation}
Detailed evaluation of this expression to yield the 3D PCTF will be discussed in Section~\ref{sec:3d_ctf}.

\subsection{Phase transfer and its correction: acBF}

Registration aligns the positions of features, but their phase contrast can still differ between the virtual images formed from each detector pixel. 
Rose's formalism gives the per-pixel PCTF \citep{rose_1977_Nonstandardimagingmethods}
\begin{equation}
\begin{aligned}
\PCTF(\boldsymbol q,\boldsymbol\Theta)
=\frac{iA(\boldsymbol\Theta)}{2\Omega_0}\Bigl\{
&A(\boldsymbol\Theta-\lambda\boldsymbol q)
e^{-i[\chi(\boldsymbol\Theta)-\chi(\boldsymbol\Theta-\lambda\boldsymbol q)]}
\\[-2pt]
-{}&A(\boldsymbol\Theta+\lambda\boldsymbol q)
e^{+i[\chi(\boldsymbol\Theta)-\chi(\boldsymbol\Theta+\lambda\boldsymbol q)]}
\Bigr\},
\end{aligned}
\label{eq:tilt_pctf}
\end{equation}
where $A$ is one inside the probe-forming aperture and zero outside,
$\Omega_0$ is its area, and $\chi$ is the aberration function.
We use the detector-resolved form of Rose's theory and the acBF correction developed in Refs.~\citep{ma_2026_Information4DSTEMWhere,ma_2026_UsingAberrationsImprove}, with $\lambda\boldsymbol q$ replacing Rose's angular scattering coordinate\footnote{Equation \ref{eq:tilt_pctf} differs from Equation 33 of Rose (1976), as the original assumed $\chi$ to be even and so is incorrect for nonround aberrations. The detailed derivation is given in Appendix A of \cite{ma_2026_UsingAberrationsImprove}}.
Here we extend that treatment to the depth dependence and the resulting three-dimensional transfer by including the $z$-dependent terms from the curvature of the Ewald sphere, which were ignored in most previous treatments (including our own), with the limits of the approximation noted and used in section 5 of Rose's 1976 paper\cite{rose_1977_Nonstandardimagingmethods}.

For defocus alone, substituting
$\chi=-k_0(\Delta f-z)|\boldsymbol\Theta|^2/2$ gives
\begin{equation}
\begin{aligned}
\PCTF(\boldsymbol q,\boldsymbol\Theta)
= {}&\frac{iA(\boldsymbol\Theta)}{2\Omega_0}
e^{i(2\pi\boldsymbol q)\cdot\boldsymbol\Theta(\Delta f-z)}
\\
&\times\Bigl[
A(\boldsymbol\Theta-\lambda\boldsymbol q)
e^{-i\pi\lambda(\Delta f-z)|\boldsymbol q|^2}
\\[-2pt]
&\hspace{2em}-A(\boldsymbol\Theta+\lambda\boldsymbol q)
e^{+i\pi\lambda(\Delta f-z)|\boldsymbol q|^2}
\Bigr].
\end{aligned}
\label{eq:fresnel_shift_factorization}
\end{equation}
The first exponential term, which factors out of the brackets, corresponds to the real-space image shift in Eq.~\eqref{eq:tcbf_fourier_registration}. 
The remaining terms are referred to as the sidebands, and modulate the information transfer of the 4D-STEM measurement. 

acBF extends tcBF by also correcting the phase factor remaining in this transfer function. Let
$\PCTF_z^{\mathrm{res}}$ denote Eq.~\eqref{eq:tilt_pctf} evaluated at the
depth $z$, after the common shift factor has been removed. Then
\begin{equation}
\begin{aligned}
\tilde J_z^{\mathrm{acBF}}(\boldsymbol q)
=\int_{|\boldsymbol\Theta|<\alpha}
&\tilde D(\boldsymbol q,\boldsymbol\Theta)
e^{-i(2\pi\boldsymbol q)\cdot\boldsymbol\Theta(\Delta f-z)}
\\[-2pt]
&\times e^{-i\arg\left[ \PCTF_z^{\mathrm{res}}(\boldsymbol q,\boldsymbol\Theta)\right]}
\dd^2\boldsymbol\Theta.
\end{aligned}
\label{eq:acbf_correction_phasor}
\end{equation}
Correcting only the phase of the per-pixel CTF avoids amplifying noise in the reconstruction.
An inverse spatial Fourier transform gives the depth
section $J_z^{\mathrm{acBF}}$.

\subsection{Two-Atom Simulation: Depth Sectioning}
\label{sec:two_atom}

To demonstrate the depth-resolving capabilities of tcBF and acBF imaging, we simulated two laterally offset carbon atoms separated by 100~\AA{} along the beam direction. 
The known atom planes are assigned to $z=0$ and $z=100$~\AA{}.
The simulation used a 300~keV probe with a 30~mrad convergence semi-angle and overfocus of $100$~\AA{} (so that the beam crossover sits $100$~\AA{} before the top atom).
The probe was scanned with a step of 0.9 times the probe Nyquist sampling, and reconstructed without upsampling. 

We reconstructed both depth section stacks from this single 4D dataset using the operations in Equations~\ref{eq:tcbf_fourier_registration} and \ref{eq:acbf_correction_phasor} for tcBF and acBF.
Both methods localize the atoms in their respective depth planes, but tcBF shows oscillatory rings around the atoms in three dimensions. 
acBF corrects the PCTF phase and more compactly concentrates the contrast near the selected atom. 
The maxima in both volumes remain elongated along the beam, reflecting the greatly limited depth resolution in comparison to the transverse resolution. 
In section~\ref{sec:3d_ctf} we derive the 3D contrast transfer function that relates this elongation to the small axial bandwidth at low lateral spatial frequency.
The acBF reconstruction also shows similar shapes for the atoms at both depths, while the appearance of the two atoms differs substantially between the two atoms in tcBF due to the lack of CTF correction.

\begin{figure}[!htbp]
    \centering
    \includegraphics[width=1.0\textwidth]{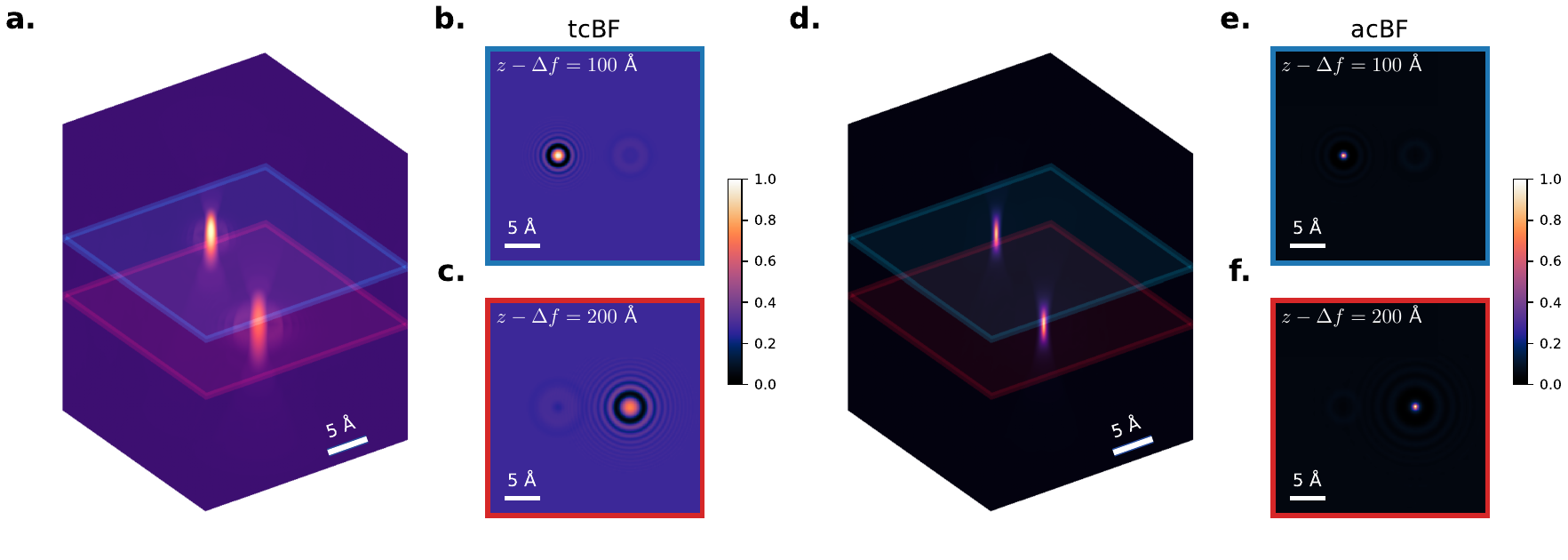}
    \caption{tcBF (a--c) and acBF (d--f) depth sectioning of two simulated carbon atoms separated by 100~\AA{} along the beam direction. 
    (a,d) Three-dimensional reconstruction volumes with the two known atom planes overlaid. 
    (b,e) Transverse sections at $z-\Delta f=100$~\AA{}. (c,f) Transverse sections at $z -\Delta f=200$~\AA{}. 
    The simulation uses 300~keV electrons and a 30~mrad convergence semi-angle. 
    The tcBF reconstruction shows oscillatory real-space tails from the PCTF and the two atoms have different appearance in their central planes, 
    whereas acBF corrects its for contrast transfer effects and concentrates the contrast both laterally and in depth with a sharp and non-oscillatory response.}
    \label{fig:two_atom_sectioning}
\end{figure}

\subsection{Depth-dependent contrast transfer}

\begin{figure}[!htbp]
    \centering
    \includegraphics[width=0.7\textwidth]{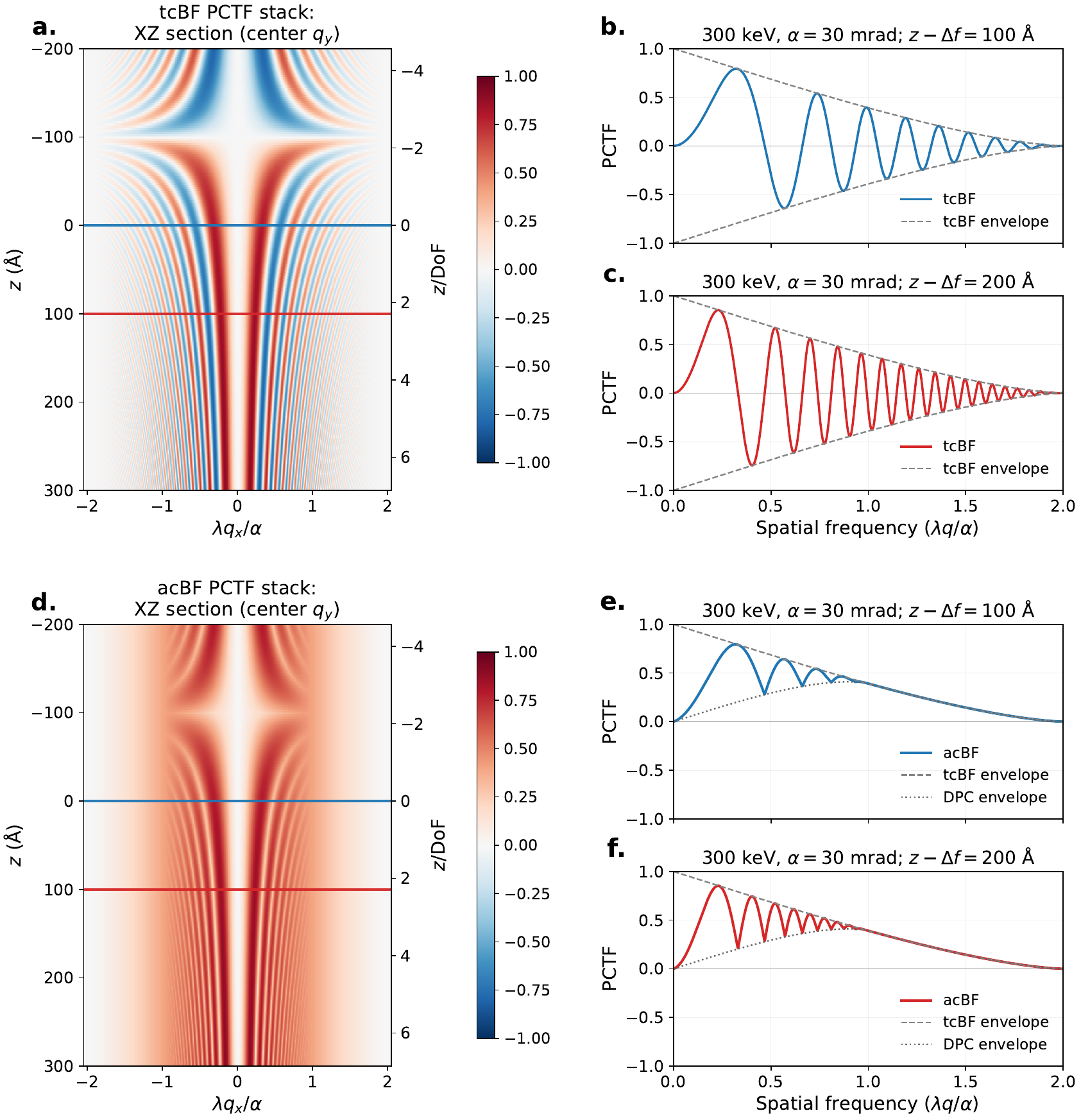}
    \caption{Stack of depth-dependent 2D PCTFs for the conditions used in the two-atom simulation 
    in Fig.~\ref{fig:two_atom_sectioning}. 
    (a) $q_x$--$z$ section through the tcBF PCTF stack. 
    The blue and red lines mark the depths of the two atoms. 
    (b,c) Radial tcBF PCTFs at the atom planes, together with the tcBF envelope. (d) 
    Corresponding section through the acBF PCTF stack. 
    (e,f) Radial acBF PCTFs at the same two depths, shown with the positive 
    tcBF and DPC envelopes. The PCTFs of both methods vary rapidly within one depth-of-field from the beam crossover, and become more constatnt at larger defocus. }
    \label{fig:two_atom_pctf_response}
\end{figure}

Figure~\ref{fig:two_atom_pctf_response} shows the corresponding 2D PCTFs evaluated at different depths. 
This is a stack of projected lateral-frequency responses, and does not give the full 3D response of an object, which we derive later in Section~\ref{sec:3d_ctf}. 
The tcBF pCTF crosses zero repeatedly as the defocus phase increases, with the locations of the zeros shifting rapidly with increasing depth around the focal plane and eventually becoming more consistent at larger distances.
Increasing the magnitude of the effective defocus from 100 to 200~\AA{} approximately doubles the number of zero crossings over the same frequency interval.

acBF combines the two sidebands at each detector pixel and corrects the phase
of their sum. 
Its transfer is therefore nonnegative at each evaluated depth.
Where both sidebands overlap, at spatial frequencies $[0,\alpha]$, double-sideband interference still modulates the PCTF magnitude, with the modulation varying with distance of the feature to the crossover, but never crossing through zero. 
This modulation varies as the feature moves away from the beam crossover plane.
For features at the crossover, only the DPC contrast transfer function is achievable, as the measurement is in-focus for features at this plane and no double-sideband information is produced.
Features further from the crossover get the benefits of increased contrast transfer induced by defocus \citep{ma_2026_Information4DSTEMWhere}.
The variation of PCTF with depth becomes more gradual for features further from the crossover, indicating that three-dimensional reconstruction is improved when operating at defocus of greater than approximately twice the optical depth of field, defined as
\begin{equation}
\mathrm{DoF}=\frac{2\lambda}{\alpha^2}.
\label{eq:probe_dof}
\end{equation}
Reduced variation in the final, corrected, PCTF between different depth sections is likely beneficial for producing an extended-depth-of-field image from the reconstructed volume, to avoid local contrast transfer changes that could make image interpretation difficult \citep{hovden_2011_ExtendedDepthField,zhang_2025_AtomByAtomMoirePhasons}.
Larger defocus values have little benefit to information transfer and require finer detector sampling to retain coherence \citep{ma_2026_UsingAberrationsImprove}; this effect similarly limits the practical sample thickness.
Where only a single sideband is present, for spatial frequencies $(\alpha, 2\alpha)$, the acBF PCTF magnitude is independent of defocus or feature depth (and this property holds for all aberrations). 
The isolated atoms simulated here ensure weak scattering behavior, while inter-plane multiple scattering in thick or crystalline also affects the response and falls outside the linear model used here.

\section{A tomographic interpretation of 4D-STEM}
\label{sec:single_atom}

The conventional approach to recovering 3D information in electron microscopy is tomography, where the object is tilted through a series of angles along one or two axes and imaged at each orientation, providing different projections of the same structure \citep{ercius_2015_ElectronTomography}. 
As the view changes, a point feature at a fixed position in the object moves across these projections, and its trajectory carries information about both its lateral position and depth. 

\begin{figure}[!htbp]
    \centering
    \includegraphics[width=1.0\textwidth]{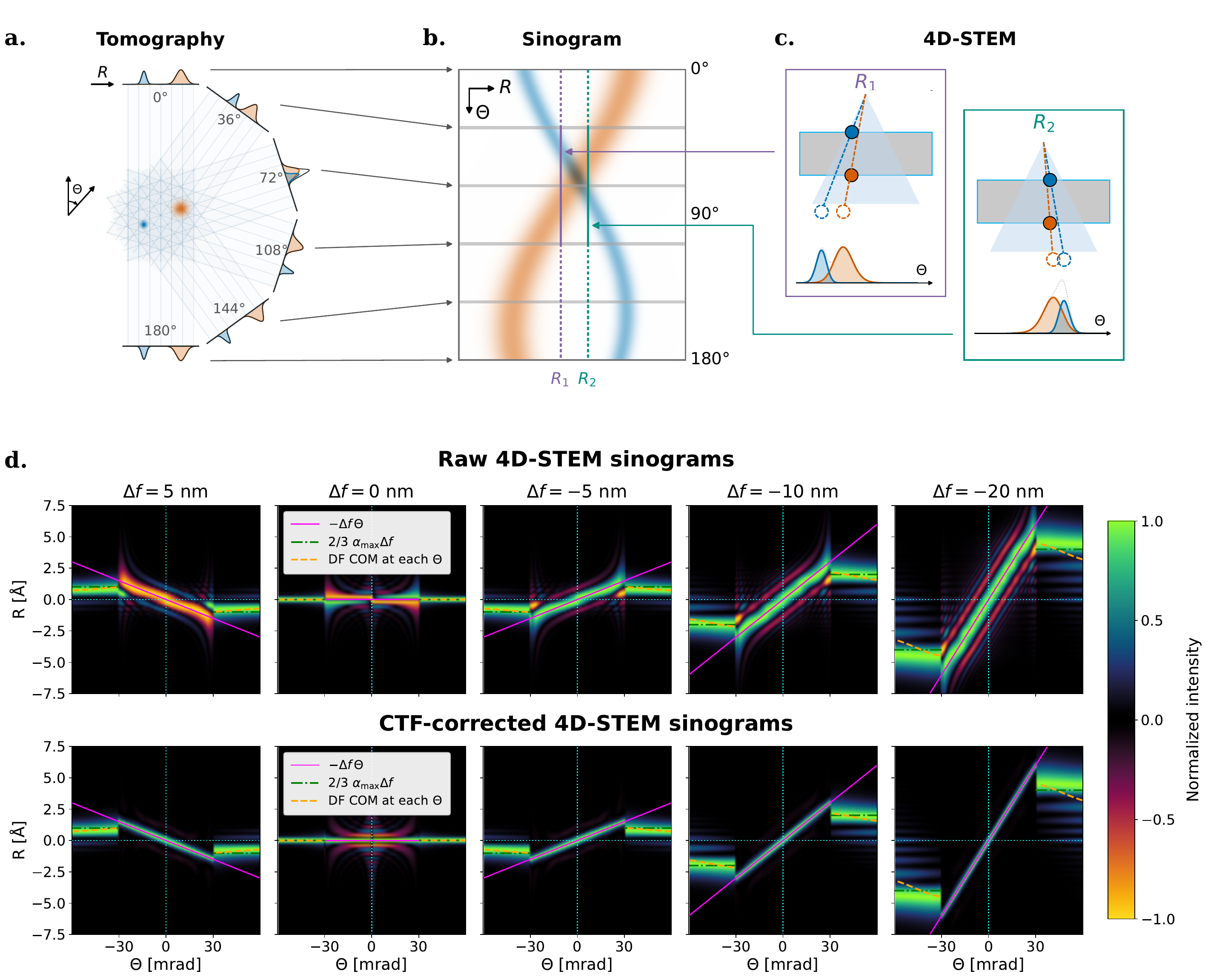}
    \caption{
    (a-c) Construction of a sinogram (b) by (a) tomography or (c) 4D-STEM. 
    By sweeping through angle $Q$ or position $R$, rays through two point objects map to 
    sinusoidal trajectories in the position--angle sinogram.
    (a) Successive projections in tomography. At each viewing angle $Q$, a profile
    is recorded over position $R$. (b) Corresponding sinogram over
    $Q=0$--$180^\circ$, with $R$ horizontal and $Q$ vertical.
    Each object produces a sinusoidal band over the angular range, blue and orange identifying
    two point objects at different positions. (c) Spatial sampling in 4D-STEM.
    Each scan position records the available angles simultaneously.
    (d) Single-atom 4D-STEM sinograms at five nominal defocus values. 
    The upper row contains raw sinograms, while 
    the lower row contains the same data after PCTF correction but without shift correction. 
    The dark-field signal ($|\Theta|>30$~mrad) is left unchanged in both, apart from being rescaled for visibility. 
    The magenta line is the defocus-only geometric trajectory 
    $r_\parallel=-\Delta f\,\Theta_x$, the green dash--dot segments mark 
    $\pm(2/3)\alpha|\Delta f|$, and the orange dashed curves are independently 
    measured dark-field centers of mass. For this defocus-only series, 
    the slope changes sign with defocus and vanishes at focus, 
    demonstrating that depth is encoded as parallax in one 4D dataset.}
    
    \label{fig:single_atom_sinograms}
\end{figure}

A sinogram displays this motion by placing line profiles from successive angular views next to one another \citep{ercius_2015_ElectronTomography, radon_1986_DeterminationFunctions}.
In parallel-beam tomography, each profile is a line-integral projection of the object at one projection angle. 
In an $xz$ section through the sample, rotation about the $y$ axis by an angle $\theta$ causes a feature at the point $(x,z)$ to appear on the detector at the position
\begin{equation}
x'=x\cos\theta-z\sin\theta,
\label{eq:tomographic_projection_coordinate}
\end{equation}
where $\theta$ is the projection angle. For an ideal projection of a volume
$f(x,z)$, the measured profile is the line integral
\begin{equation}
p_\theta(x')=\int_{-\infty}^{\infty}
f\!\left(x'\cos\theta+\tau\sin\theta,
-x'\sin\theta+\tau\cos\theta\right)\dd\tau,
\label{eq:tomographic_line_integral}
\end{equation}
where $\tau$ measures distance along the beam.
These projections are recorded over a range of $\theta$ and stacking these projections gives the sinogram, or mathematically the Radon transform \citep{bracewell_1956_StripIntegrationRadioAstronomy,kak_2001_PrinciplesComputerizedTomographicImaging}.
Thus, one axis is position along the profile and the other is viewing angle.
By Equation~\ref{eq:tomographic_projection_coordinate}, a point in the object traces a sinusoid as the viewing direction rotates, giving the plot its name (Fig.~\ref{fig:single_atom_sinograms}a,b).
Sinograms are widely used in tomography data processing to evaluate the quality of axis alignment, as geometric inconsistencies \citep{ercius_2015_ElectronTomography} cause the trace to deviate from the ideal sinusoidal shape.

An equivalent procedure can be used to construct sinograms from a 4D-STEM dataset.
To do so, we take a line through the virtual images formed from each detector pixel along single row. 
This gives a position--angle section of the 4D data without rotating the specimen, instead using the reciprocity between detector position and beam tilt to produce the different views.  
Since the probe aperture spans only a small angular range, the angular range of the sinogram is correspondingly narrow (typically less than $\pm2^\circ$, though potentially higher with next-generation aberration correctors \citep{ishikawa_2021_AutomatedGeometricAberrationCorrection,uhlemann_2022_LargeApertureSTEMHexapole,koibuchi_2024_HighResolutionSTEMTiltedSpecimen}).
As a result, defocus produces an approximately straight trajectory whose slope measures distance from the crossover, rather than showing a full sinusoidal curve. 
Unlike in conventional tomography, where the imaging system stays stationary and the sample is rotated, the contrast along a trajectory also changes with detector angle because each virtual image has its own PCTF \citep{ren_2020_MultipleScatteringTomography}. 
The two acquisition strategies also fill this position--angle array in different orders: tomography records a spatial projection at each angle, whereas 4D-STEM records all detector angles at each scan position (Fig.~\ref{fig:single_atom_sinograms}c).

Tomographic reconstruction combines these various projections to determine the 3D structure of the original object.
For ideal line-integral projections, the unfiltered backprojection algorithm evaluates each profile at the position given by Eq.~\eqref{eq:tomographic_projection_coordinate} and sums over angles:
\begin{equation}
\begin{aligned}
b(x,z)
&=\sum_\theta p_\theta(x\cos\theta-z\sin\theta)\\
&\simeq\sum_\theta p_\theta(x-z\theta)
\qquad (|\theta|\ll 1).
\end{aligned}
\label{eq:sinogram_small_angle_backprojection}
\end{equation}
Applying a small-angle approximation to this approach reduces it to a summation of parallax-shifted copies of each projection.
This is the geometric analogue of tcBF registration in Eq.~\eqref{eq:realspace_projection}, by replacing $z$ in the tomography convention with $\Delta f-z$, as the effective pivot point for tilting of the 4D-STEM images is the probe crossover.
However, the virtual BF images are not ideal projection images due to the phase contrast image formation mechanism, which also varies between the different views.

\subsection{Single-atom sinograms}
The sinogram construction introduced in Section~\ref{sec:virtual_images_sinograms} lets us demonstrate these two distinct effects: the overall shifting trajectory produced by the parallax effect, versus the fringes and sign reversals along it that result from wave-optical effects. 
We use a simulated 4D-STEM dataset of a single isolated carbon atom so that neither effect is obscured by overlapping features.
We simulated an isolated carbon atom plated at $(50,50,5)$~\AA{} in a $100\times100\times10$~\AA{} cell, with a projected potential sampling of 0.05~\AA{}.
A 300~keV probe with a 30~mrad convergence semi-angle and a 1~mrad aperture rolloff was scanned over a $30\times30$~\AA{}$^2$ field centered on the atom.
The scan step was 0.9 times the probe Nyquist sampling.
Diffraction patterns were recorded to 60~mrad and sum-binned by four, giving $204\times204\times152\times152$ samples in scan and detector coordinates.

We calculated five datasets at defocus values of
$+5$, $0$, $-5$, $-10$, and $-20$~nm, with respect to the position of the atom.
For each dataset, we extracted a line through the center of the field of view from each virtual image along the central detector row. 
Arranging these profiles against $\Theta_x$ produces the sinograms in Fig.~\ref{fig:single_atom_sinograms}d.

In the detector-axis orientation used in Fig.~\ref{fig:single_atom_sinograms},
the expected trajectory is
\begin{equation}
r_\parallel(\Theta_x)=r_{\parallel,0}-\Delta f\,\Theta_x,
\qquad
\frac{\partial r_\parallel}{\partial\Theta_x}=-\Delta f,
\label{eq:single_atom_parallax_slope}
\end{equation}
where $r_\parallel$ is position along the profile and $r_{\parallel,0}$ is the on-axis atom position. 
The simulated trajectories are horizontal at focus, reverse slope across
focus, and become steeper as $|\Delta f|$ increases.
The shift-correction step of tcBF reconstruction amounts to a shearing of these sinograms before summation over the angular coordinate. 

The fringes around each trajectory are produced by the contrast transfer functions for each off-axis image, and their spacing decreases with increasing defocus.

\subsection{Aberration-corrected sinograms}

To highlight the impact of the phase correction step of acBF reconstruction, we corrected the defocus-dependent PCTF phase inside the BF disk without applying the shift correction, producing the sinograms shown in the lower row of Fig.~\ref{fig:single_atom_sinograms}d.
The phase correction nearly eliminates the oscillations around the central trajectory, yielding a single sharp line across the bright-field disk.
The contrast reversal for $\Delta f = 5$~nm case, where the atom sits above the crossover point, has also been eliminated. 
For aberrations beyond defocus, the appearance of the sinogram is more complicated; this additional complexity, and its consequences for reconstruction, are discussed in Section~\ref{sec:higher_order_depth}

In both the raw and corrected sinograms, pixels outside the bright-field disk were left unmodified, as they do not posses a linear PCTF \citep{ma_2026_Information4DSTEMWhere}.
However, the differently-sloped trajectories of the dark-field region between the different defocus values shows that this portion of the detected signal also includes parallax information.
The shifts for the dark-field images are roughly given by $(2/3)\alpha\Delta f$, though this approximation is not universal and the true shifts depend on both the sample and the defocus in a complicated manner \citep{ma_2026_Information4DSTEMWhere}. 
Bright-field phase images and tilt-corrected dark-field images have also been combined experimentally using non-iterative Fourier-space fusion weighted by their spectral signal-to-noise ratios \citep{you_2026_GapFreeInformationTransfer}.
This combines complementary contrast channels, but does not make the dark-field signal linear in the specimen phase.

\section{A light-field interpretation of 4D-STEM}
\label{sec:lf_discussion}

\begin{figure}[!htbp]
    \centering
    \includegraphics[width=1.0\textwidth]{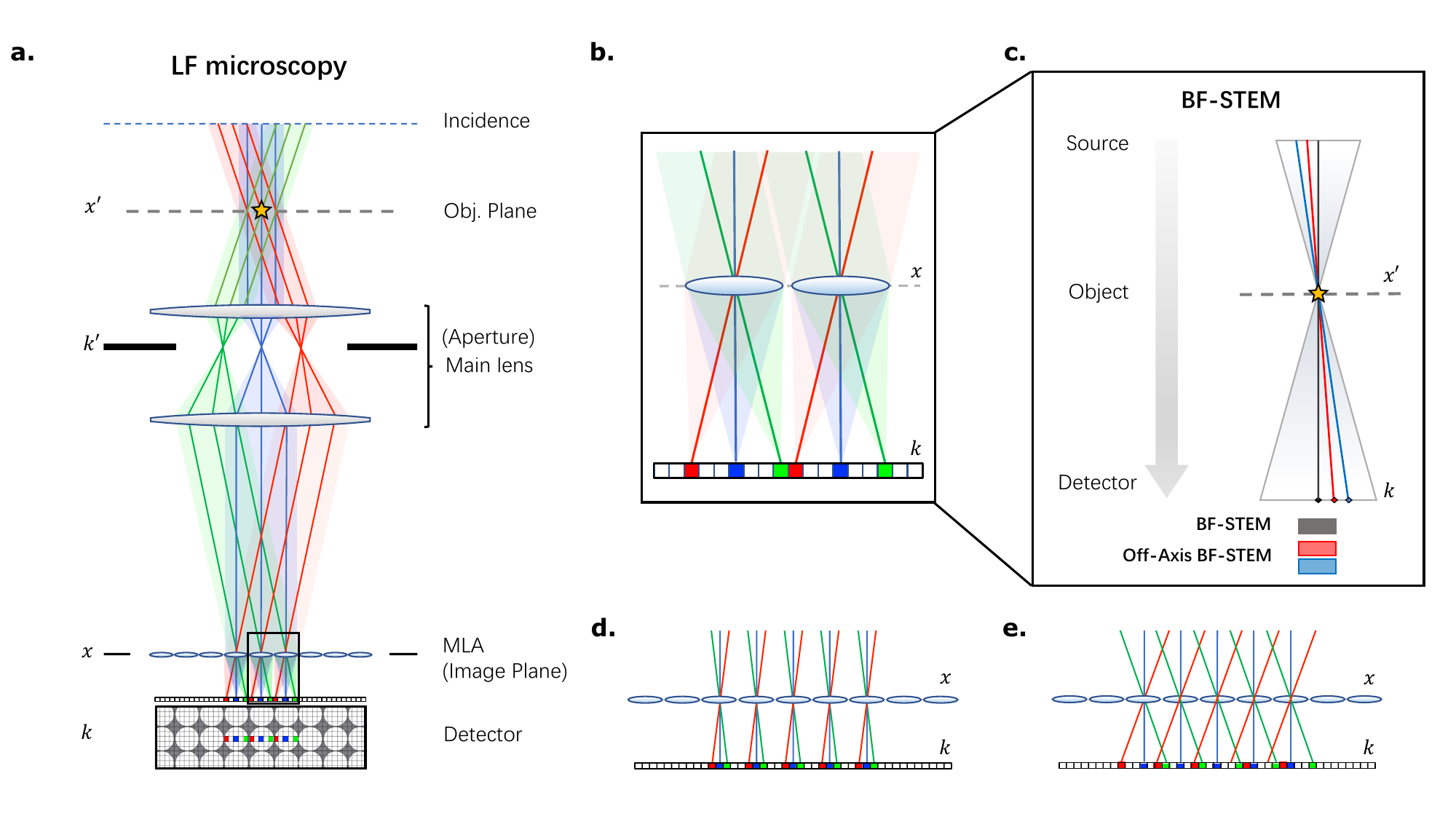}
    \caption{Optical light-field imaging and its reciprocal 4D-STEM geometry.
    (a) A microlens array records position and propagation direction at the
    image plane of a light-field microscope. (b) Spatial samples are resolved
    into angular samples on the detector. (c) In BF-STEM, an on-axis or
    off-axis detector pixel is reciprocal to a corresponding illumination
    direction in conventional TEM. (d,e) The maximum ray angle which can be recorded with a microlens array is limited by crosstalk between adjacent lenslets, which produces an aliasing artifact when the high angle rays entering one lenslet lands on the detector in a pixel belonging to its neighbor (e). In 4D-STEM, each diffraction pattern is recorded separately at
    successive scan positions, so this spatial--angular mixing does not occur; the
    angular range is instead set only by the camera length and detector extent.}
    \label{fig:lf_correspondence}
\end{figure}

A closely related position--angle concept exists in optical imaging in the form of the light-field, or plenoptic, camera. 
The light field of a scene $L(x,y,u,v)$ is the intensity of the ray of light that passes through the position $(u,v)$ at the lens plane and arrives at $(x,y)$ at another, further, plane in the system. 
By relating the scattering angle to the position in the lens plane, we immediately recognize that this is a position--angle representation of the distribution of scattered light.
In this sense, it is closely related to both the eikonal function of geometric optics \citep{rose_2012_GeometricalChargedParticleOptics} and the Wigner distribution in wave optics \citep{ma_2025_Emittanceminimizationaberration,ma_2025_Emittanceminimizationaberrationb}.
A conventional camera records only the intensity distribution as a function of $(x,y)$, with no sensitivity to the ray angles (apart from the effect of the limiting aperture). 
In light-field photography (or microscopy), a mixed position--angle recording is produced by placing a microlens array at a plane in front of the sensor.
Each microlens selects a spatial region in the object plane, and the detector pixels beneath it distinguish ray directions \citep{ng_2005_LightFieldPhotography,ng_2006_Digitallightfield}.
This setup is shown schematically in Figure~\ref{fig:lf_correspondence}a for the light-field microscope, including the rays for normal and tilted illumination. 
Thus, both angular and position information are simultaneously encoded in the sensor plane. 

Selecting the same angular sample under every microlens gives a view through one part of the main lens aperture. 
Features at different depths move by different amounts between these views, by the same parallax mechanism as discussed previously. 
Registering and summing the views therefore refocuses the image after acquisition, using a single recorded light field rather than a focal series \citep{ng_2006_Digitallightfield}.
This post-exposure refocusing capability has been included in a consumer light-field camera \citep{cho_2013_LytroCalibration} as well as used in laboratory microscopy \citep{bimber_2019_LightFieldMicroscopyReview}.

\begin{figure}[!htbp]
    \centering
    \includegraphics[width=\textwidth,keepaspectratio]{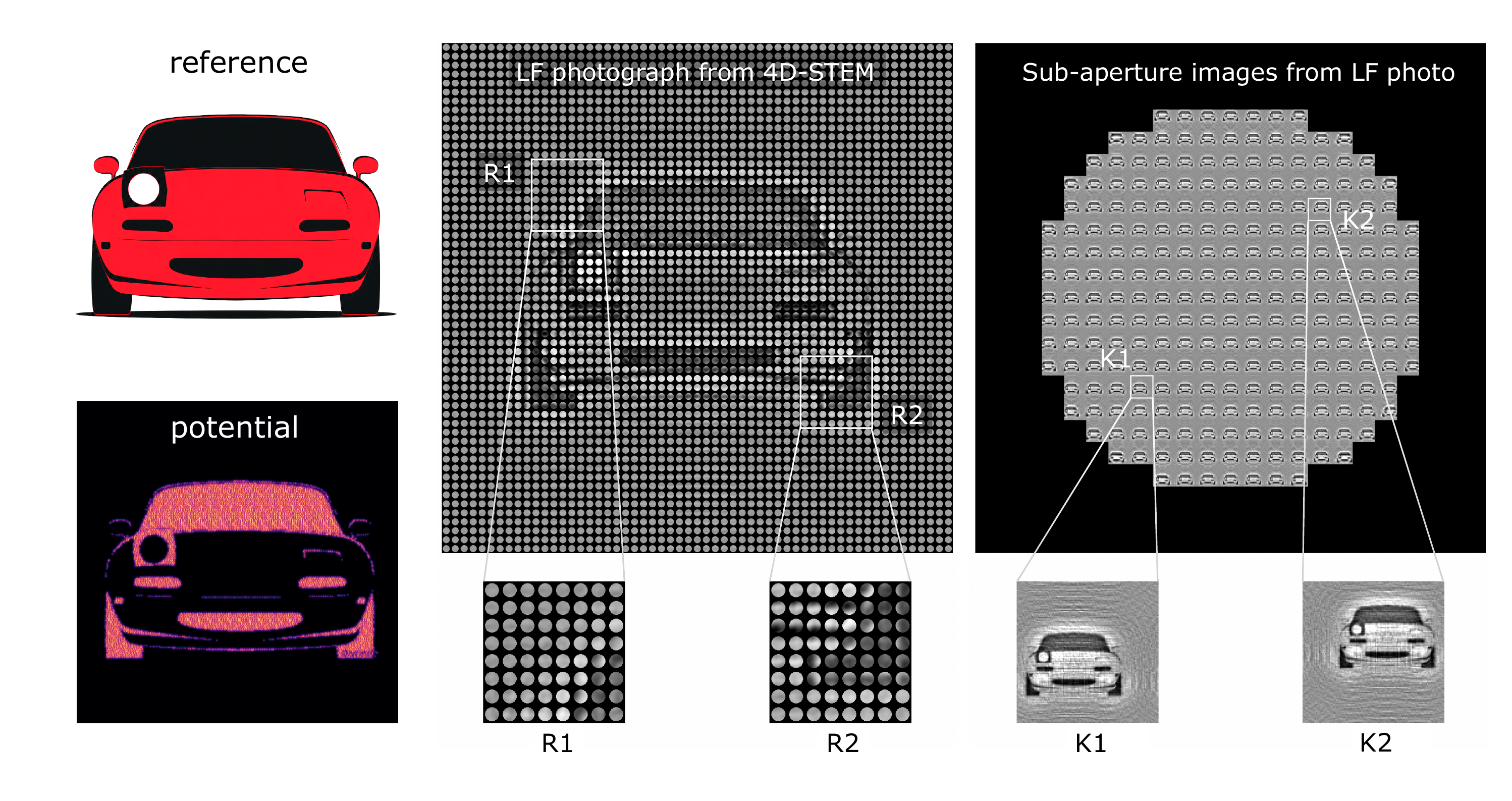}
    \caption{Rearrangement of a simulated 4D-STEM dataset as a light-field photograph. A carbon test object shaped as a car is scanned to form the position--angle mosaic, with one diffraction pattern at each scan position. The bottom panels magnify the marked scan regions $R_1$ and $R_2$, each containing multiple diffraction patterns. Alternatively, sampling a fixed coordinate within every diffraction pattern gives a detector-resolved mosaic, as shown in the right panel with two selected views enlarged below. The two arrangements contain the same four-dimensional samples and expose complementary spatial and angular views.}
    \label{fig:lf_miata}
\end{figure}

There is currently no direct analog to the microlens array in electron microscopy.
The crossed laser phase plate \citep{petrov_2026_CrossedLaserPhasePlates} could, potentially, be used as an approximation to such a device by altering the setup of the coupling optics to place an image plane at the phase grating produced by the laser standing wave rather than the diffraction plane, albeit with rather poor focusing properties of the ``lenslets.''
However, we can view the position--angle nature of a 4D-STEM acquisition as a serially-acquired version of the light-field photograph. 
This correspondence is made clear by arranging the diffraction patterns from a 4D-STEM dataset on a grid corresponding to their scan position, which was in fact common in earlier 4D-STEM literature as a device to highlight the high dimensional nature of the collected data \citep{yang_2017_ElectronPtychographicPhaseImaging,ophus_2019_FourDimensionalScanningTransmission}, but the connection to lightfield imaging appears to have been neglicted until now.
This correspondence is made clear by arranging the diffraction patterns from a 4D-STEM dataset on a grid corresponding to their scan position, as in Figure~\ref{fig:lf_miata}. 
The central mosaic contains a small diffraction pattern at each scan position; $R_1$ and $R_2$ magnify two groups of these patterns. 
Selecting one detector coordinate across the entire scan instead gives one car image in the right-hand mosaic. 
Such arrangement was in fact common in earlier 4D-STEM literature, as a device to highlight the high dimensional nature of the collected data \citep{yang_2017_ElectronPtychographicPhaseImaging,ophus_2019_FourDimensionalScanningTransmission}, but the connection to lightfield imaging appears to have been neglicted until now.

A crucial difference concerns the accessible angular range, and in this case it is 4D-STEM that has an angular sampling advantage. 
In a physical LF camera, each microlens maps one spatial sample onto a finite microimage on the shared detector. 
The microlens pitch and its distance from the detector plane therefore limit the ray angles that can remain inside that spatial sample's allocated footprint. 
For rays arriving at larger angles, neighboring microimages overlap, so angular information from adjacent $(x,y)$ coordinates is mixed and produces severe spatial--angular aliasing, as illustrated by the geometry in Fig.~\ref{fig:lf_correspondence}(d,e) \citep{ng_2006_Digitallightfield,
dansereau_2013_DecodingCalibrationRectification}. 
This aliasing condition is typically avoided by choosing the limiting aperture so as to preclude any overlapping rays from entering the microlens plane.
4D-STEM does not have this microlens-overlap limitation, as the complete diffraction pattern is acquired and stored separately at every probe position, so a large detector angle cannot spill into the diffraction pattern of an adjacent scan coordinate, and so the detector sampling can be chosen freely.
This separation removes the LF-camera cross-talk mechanism, although scan-step and detector-pixel sampling can still impose their own Nyquist and aliasing limits.
The parallel nature of lightfield acquisition is beneficial for fast acquisition speed, but conversely limits the number of pixels in the measurement to the number of pixels on the detector. 
Serial acquisition of diffraction patterns in 4D-STEM means the total pixel count is in theory limited only by drift concerns; a $128\times128$ pixel scan with a $128\times128$ pixel detector (which is acquired in around 1.6~seconds with a modern direct electron detector \cite{philipp_2022_VeryHighDynamicRange}), yields a $\approx 268$~Mpix lightfield ``photograph.''


\subsection{Fourier-slice refocusing with a 4D FFT}
\label{sec:fourier_slice_refocusing}
We can, as in the case of tomography, extend the analogy to the reconstruction methods as well.
Equation~\eqref{eq:realspace_projection} is the electron-microscopy counterpart of optical shift-and-add refocusing for lightfield photography \citep{ng_2005_LightFieldPhotography,ng_2006_Digitallightfield, wu_2017_LightFieldImage,zhou_2021_Reviewlightfield}.
However, the shift-and-sum approach is not the dominant algorithm used in lightfield imaging, as there exists an efficient Fourier-based reconstruction method, which we describe below. 

The Fourier-slice theorem states that the Fourier transform of a projection is equivalent to a slice through the higher-dimensional Fourier transform by a plane through the origin and perpendicular to the projection direction. 
In the case of 3D reconstruction, the projection is the shifted angular sum in Eq.~\eqref{eq:realspace_projection} (or, equivalently, the integration along a chosen line in the sinogram), with the depth $\Delta f - z$ setting the projection direction, just as registering angular views selects a refocused plane in light-field photography \citep{ng_2005_Fourierslicephotography}.

To see the connection explicitly, extend $D$ (Eq.~\eqref{eq:detector_resolved_contrast}) by zero outside the BF aperture and Fourier transform over detector angle as well as probe position:
\begin{equation}
\mathcal F_{\mathbf r,\boldsymbol\Theta}[D](\boldsymbol q,\boldsymbol\xi)
=\int \tilde D(\boldsymbol q,\boldsymbol\Theta)
e^{-i\boldsymbol\xi\cdot\boldsymbol\Theta}\dd^2\boldsymbol\Theta.
\label{eq:fourier_4d_definition}
\end{equation}
Here $\boldsymbol\xi$ is conjugate to detector angle and $\boldsymbol q$ remains the spatial frequency in the image space. 
Comparing this expression with Eq.~\eqref{eq:tcbf_fourier_registration} gives
\begin{equation}
\tilde J_z^{\mathrm{tcBF}}(\boldsymbol q)
=\mathcal F_{\mathbf r,\boldsymbol\Theta}[D]
\!\left(\boldsymbol q,(2\pi\boldsymbol q)(\Delta f-z)\right).
\label{eq:fourier_slice}
\end{equation}
The angular integration kernels in Eqs.~\eqref{eq:tcbf_fourier_registration} and \eqref{eq:fourier_4d_definition} are identical when $\xi_x=2\pi q_x(\Delta f-z)$ and $\xi_y=2\pi q_y(\Delta f-z)$.
Equation~\eqref{eq:fourier_slice} thus turns the shift-and-sum operation in tcBF to slicing the 4D-Fourier transform of the dataset followed by a 2D Fourier transform to return to real space.

We have implemented the 4D-FFT method for tcBF reconstruction of 4D-STEM data.
This algorithm has a substantial one-time setup cost from evaluation of the 4D-FFT of the gigabyte-scale input dataset.
However, the slicing and single 2D FFT required for each depth section evaluation is considerably faster than the shift-and-sum approach, which typically requires one FFT per pixel in the BF disk (using Fourier shifting to perform the subpixel registration), repeated at every slice.
While the 4D-FFT algorithm is mathematically equivalent to conventional tcBF, slicing the tilted planes does generally require interpolation, leading to slight numerical differences in practice.

We tested the reconstruction times for both the 4D-FFT and shift-and-sum algorithms, as shown in Figure~\ref{fig:lf_tcbf_benchmark}.
The incremental time to evaluate an additional slice (neglecting the setup time for both methods) shown in Figure~\ref{fig:lf_tcbf_benchmark}a,b shows a substantial speedup of over 10$\times$ for the 4D-FFT approach for all the dataset sizes we considered.
The large fixed cost of the initial 4D FFT can make LF slicing slower for a small number of sections, but the lower incremental cost becomes advantageous as more depths are requested, as shown in Figure~\ref{fig:lf_tcbf_benchmark}c,d. 
The crossover depends on detector sampling and scan size, but in our benchmarking experiment we found that under typical conditions the 4D-FFT algorithm is faster when evaluating more than 10 slices. 
The detailed benchmark protocol is described in \ref{appendix_D_benchmark}.

The 4D-FFT algorithm corresponds directly to tcBF's geometric refocusing. 
The acBF phase correction depends on both detector angle and depth and must be included before the angular sum; it is not a single depth-independent filter of the same 4D transform. 
Wave-optical light-field reconstruction provides a related, but not identical, correction, and is generally evaluated iteratively \citep{broxton_2013_Waveopticstheory}. 
This accounts for effects beyond the ray-optical model, whereas acBF corrects the coherent phase-object transfer of each electron virtual image. 
Higher-order aberrations also require the more general registration discussed next, as the projections must be evaluated along a nonlinear trajectory.

\section{3D contrast transfer functions}
\label{sec:3d_ctf}
Having explored the connections between tcBF/acBF imaging and other 3D imaging techniques, we now seek to establish the limits on 3D information transfer by 4D-STEM. 
To do so, we start from the general detector-pixel-resolved 3D PCTF derived from Rose's coherent phase term, with the Ewald sphere phase term retained \citep{rose_1977_Nonstandardimagingmethods}.
The full derivation is given in \ref{appendix_A_rose_pctf}, which we summarize briefly here.

For lateral spatial frequency $\boldsymbol q$, axial spatial frequency $q_z$,
and detector angle $\boldsymbol\Theta$, the 3D PCTF for a single virtual image is obtained by Fourier transforming Rose's coherent phase PCTF expression with respect to $z$ yielding
\begin{equation}
\resizebox{\linewidth}{!}{$
\begin{aligned}
\PCTF_{\mathrm{3D}}
(\boldsymbol q,q_z;\boldsymbol\Theta)
=
\frac{i}{2\Omega_0}
A(\boldsymbol\Theta)
\Big[
&
A(\boldsymbol\Theta-\lambda\boldsymbol q)
e^{-i[\chi(\boldsymbol\Theta)-\chi(\boldsymbol\Theta-\lambda\boldsymbol q)]}
\delta(q_z-Q_-)
\\
&-
A(\boldsymbol\Theta+\lambda\boldsymbol q)
e^{+i[\chi(\boldsymbol\Theta)-\chi(\boldsymbol\Theta+\lambda\boldsymbol q)]}
\delta(q_z-Q_+)
\Big],
\end{aligned}
$}
\label{eq:main_3d_pctf_general}
\end{equation}
where
\begin{equation}
Q_\pm(\boldsymbol\Theta,\boldsymbol q)
=\boldsymbol\Theta\cdot\boldsymbol q
\pm\frac{\lambda}{2}|\boldsymbol q|^2
\label{eq:main_ewald_sheets}
\end{equation}
are the axial components of the spatial frequencies transferred by the detector pixel under the paraxial approximation. 
$A$ is the probe aperture, $\Omega_0$ its area, and $\chi$ the aberration function, as in Eq.~\eqref{eq:tilt_pctf}.
Each delta function places the sideband on a surface in reciprocal space termed an Ewald sheet, while the exponential gives the phase modulation on the sheet and the aperture terms determine whether each sideband is present.
To obtain the final 3D PCTF, we first apply the reconstruction corrections to this PCTF at each detector pixel (shift cancellation for tcBF, and additionally phase correction for acBF) before integrating over detector angle. 

\subsection{Analytic tcBF transfer in 3D}
\label{sec:analytic_tcbf_transfer}

To obtain the response of a tcBF depth stack, we first return to the detector-pixel signal and then apply the depth-dependent shift before summing over detector angles.
For defocus alone, $\chi(\boldsymbol\Theta) =-\pi\Delta f|\boldsymbol\Theta|^2/\lambda$, so the two remaining sideband phase factors are $\exp(i2\pi\Delta f Q_\pm)$.

The tcBF shifting multiplies both terms by
$\exp[-i2\pi(\Delta f-z)\boldsymbol\Theta\cdot\boldsymbol q]$, where $z$ is the reconstruction depth.
Substituting $Q_\pm$ from Eq.~\eqref{eq:main_ewald_sheets}, and Fourier transforming with respect to $z$  gives
\begin{equation}
\begin{aligned}
\PCTF_{\mathrm{tcBF}}^{\mathrm{3D}}(\boldsymbol q,q_z)
={}&\frac{i}{2\Omega_0}\int\dd^2\boldsymbol\Theta\,
A(\boldsymbol\Theta)
\delta(q_z-\boldsymbol\Theta\cdot\boldsymbol q)\\
&\times\Big[
A(\boldsymbol\Theta-\lambda\boldsymbol q)
e^{-i\pi\lambda\Delta f|\boldsymbol q|^2}\\
&\qquad{}-A(\boldsymbol\Theta+\lambda\boldsymbol q)
e^{i\pi\lambda\Delta f|\boldsymbol q|^2}\Big].
\end{aligned}
\label{eq:main_tcbf_mapping}
\end{equation}

Both terms lie on the same Ewald sheet $q_z=\boldsymbol\Theta\cdot\boldsymbol q$, but retain their opposite quadratic phases.

We define
\begin{equation}
\begin{aligned}
\ell_\pm(t)
&=2\sqrt{\max\!\left[0,\min\!\left\{
\alpha^2-t^2,\,\alpha^2-(t\pm\lambda|\boldsymbol q|)^2
\right\}\right]},\\
\ell_0(t)&=\min\{\ell_-(t),\ell_+(t)\}.
\end{aligned}
\label{eq:aperture_chord_lengths}
\end{equation}
which have a geometric meaning as the lengths of the intersections of the sideband regions for different detector positions.
After evaluating integral over $\boldsymbol\Theta$, this allows us to compactly write the final 3D PCTF as
\begin{equation}
\begin{aligned}
\PCTF_{\mathrm{tcBF}}^{\mathrm{3D}}(\boldsymbol q,q_z)
=\frac{i}{2\Omega_0|\boldsymbol q|}\Big[&
\ell_-\!\left(\frac{q_z}{|\boldsymbol q|}\right)
e^{-i\pi\lambda\Delta f|\boldsymbol q|^2}\\
&-\ell_+\!\left(\frac{q_z}{|\boldsymbol q|}\right)
e^{i\pi\lambda\Delta f|\boldsymbol q|^2}\Big].
\end{aligned}
\label{eq:tcbf_3d_closed_form}
\end{equation}
Note that at $\boldsymbol q=0$, the first-order phase transfer vanishes and expressions containing $1/|\boldsymbol q|$ are not evaluated there.

In the $q_y=0$ section, for $0<q_x<2\alpha/\lambda$, the two tcBF branches occupy
\begin{equation}
\begin{aligned}
-\alpha q_x+\lambda q_x^2&<q_z<\alpha q_x,\\
-\alpha q_x&<q_z<\alpha q_x-\lambda q_x^2.
\end{aligned}
\label{eq:main_tcbf_boundaries}
\end{equation}
Their union broadens with lateral frequency and splits into two separated bands when $q_x>\alpha/\lambda$.
Negative lateral frequencies follow by inversion symmetry.

We show slices of this 3D transfer function, both by numerical FFT of a reconstructed test object in Figure~\ref{fig:single_atom_analytic_ctf}a and from the theoretical expression in Figure~\ref{fig:tcbf_acbf_3d_ctf_boundaries}a,c. 
The numerical result shows excellent agreement with the theory. 
Inspection of the complex PCTF reveals that it is conjugate symmetric, and so corresponds to a purely real point spread function (PSF) in real space.
The plots are displayed in units (detailed in \ref{appendix_B}) which have been normalized by the diffraction limit for transverse spatial frequencies and by the depth of field for axial frequencies.
The takeoff angles of the envelope function are thus the probe semi-convergence angle $\alpha$. 

\subsection{Analytic acBF transfer in 3D}
\label{sec:analytic_acbf_transfer}

acBF starts from the same two detector-pixel sideband terms, but supplements the tcBF shift with the residual phase correction in Eq.~\eqref{eq:acbf_correction_phasor}.
This correction is evaluated based on the effective defocus at the reconstruction plane and so varies with reconstruction depth (as do the shifts).
At each selected depth, the correction is evaluated at the corresponding effective defocus and applied to the full 3D sideband expression, retaining the Ewald sheet delta functions (Eq.~\eqref{eq:acbf_3d}).
In a single-sideband region, only one shifted aperture term produces contrast and the transfer magnitude is independent of defocus.
In a double-sideband region, both overlap and the magnitude contains
\begin{equation}
\left|e^{-i\pi\lambda\Delta f|\boldsymbol q|^2}
-e^{+i\pi\lambda\Delta f|\boldsymbol q|^2}\right|
=2\left|\sin\!\left(\pi\lambda\Delta f|\boldsymbol q|^2\right)\right|.
\label{eq:main_dsb_defocus_oscillation}
\end{equation}
This interference accounts for the residual defocus modulation of the acBF curves in Fig.~\ref{fig:two_atom_pctf_response}.

For the same uniform circular aperture and defocus-only aberration, the Ewald sheets select
$t_\pm=(q_z\mp\lambda|\boldsymbol q|^2/2)/|\boldsymbol q|$, and
\begin{equation}
\begin{aligned}
&\PCTF_{\mathrm{acBF}}^{\mathrm{3D}}(\boldsymbol q,q_z)
=\frac{1}{2\Omega_0|\boldsymbol q|}\Big\{
\ell_-(t_-)+\ell_+(t_+)-\ell_0(t_-)-\ell_0(t_+)\\
&\quad{}+i\operatorname{sgn}\!\left[\sin(\pi\lambda\Delta f|\boldsymbol q|^2)\right]
\Big[\ell_0(t_-)e^{-i\pi\lambda\Delta f|\boldsymbol q|^2}
-\ell_0(t_+)e^{i\pi\lambda\Delta f|\boldsymbol q|^2}\Big]\Big\}.
\end{aligned}
\label{eq:acbf_3d_closed_form}
\end{equation}
The first line contains the defocus-independent single-sideband contribution, and the remaining terms contain the double-sideband contributions which depend on the defocus.
At zeroes of the $\sin$ term we suppress that contribution, taking $\operatorname{sgn}(0)=0$.
Integrating the complex expressions in Eqs.~\eqref{eq:tcbf_3d_closed_form} and \eqref{eq:acbf_3d_closed_form} over $q_z$ recovers the tcBF and acBF 2D transfer functions in Eqs.~\eqref{eq:tcbf_2d_collapse} and \eqref{eq:acbf_2d_collapse}.

Mapping the aperture overlaps onto the Ewald sheets gives the envelope
\begin{equation}
-\alpha|q_x|+\frac{\lambda q_x^2}{2}
<q_z<
\alpha|q_x|-\frac{\lambda q_x^2}{2},
\qquad |q_x|<2\alpha/\lambda.
\label{eq:main_acbf_full_support}
\end{equation}
Unlike the tcBF case which branches out at higher spatial frequencies, this envelope closes at the lateral cutoff.
The DSB part requires simultaneous overlap of $A(\boldsymbol\Theta)$, $A(\boldsymbol\Theta-\lambda\boldsymbol q)$, and $A(\boldsymbol\Theta+\lambda\boldsymbol q)$, and exists only for
$|q_x|<\alpha/\lambda$ in this section. 
The same information transfer envelope has been identified by Yang et al \cite{yang_2016_SimultaneousAtomicResolution} for Wigner distribution deconvolution ptychography and is equivalent to the envelope found in Terzoudis-Lumsden et al for phase scanning confocal microscopy with collection angle equal to the convergence angle \cite{terzoudis-lumsden_2023_ResolutionVirtualDepth}.

\begin{figure}[!htbp]
    \centering
    \includegraphics[width=\textwidth]{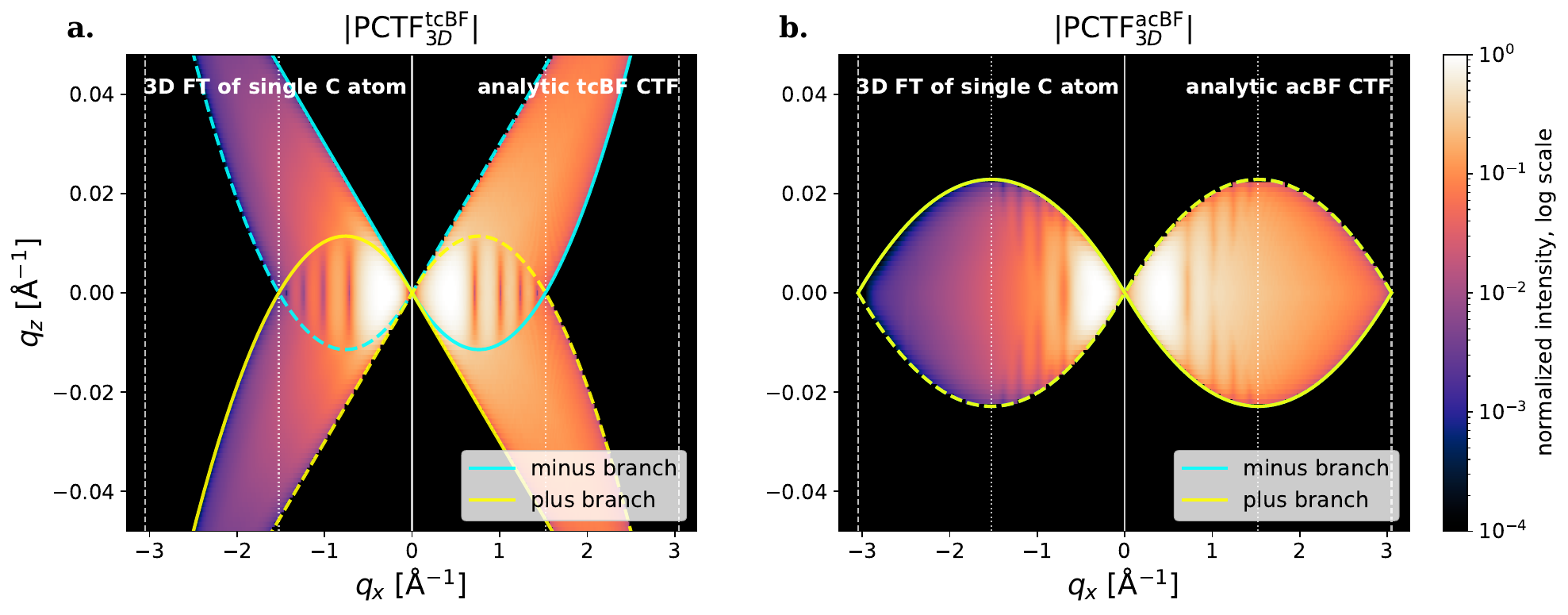}
    \caption{Numerical and analytical validation of the 3D PCTFs
    at 10~nm defocus for 300~kV electrons and $\alpha=30$~mrad for (a) tcBF and (b) acBF. In each panel, the left half is the normalized intensity of the 
    3D Fourier transform of a simulated single-atom reconstruction, 
    and the right half is the corresponding analytical PCTF intensity. 
    Solid and dashed curves mark the two analytic 
    sideband-support boundaries, while vertical guides indicate the convergence-aperture 
    frequencies. The numerical response follows the predicted broad tcBF wedge 
    and the closed acBF envelopes. A filter was used to suppress numerical instabilities near zero frequency in the simulation. Intensities are displayed on a 
    common logarithmic scale.}
    \label{fig:single_atom_analytic_ctf}
\end{figure}

\begin{figure}[!htbp]
    \centering
    \includegraphics[width=0.9\textwidth]{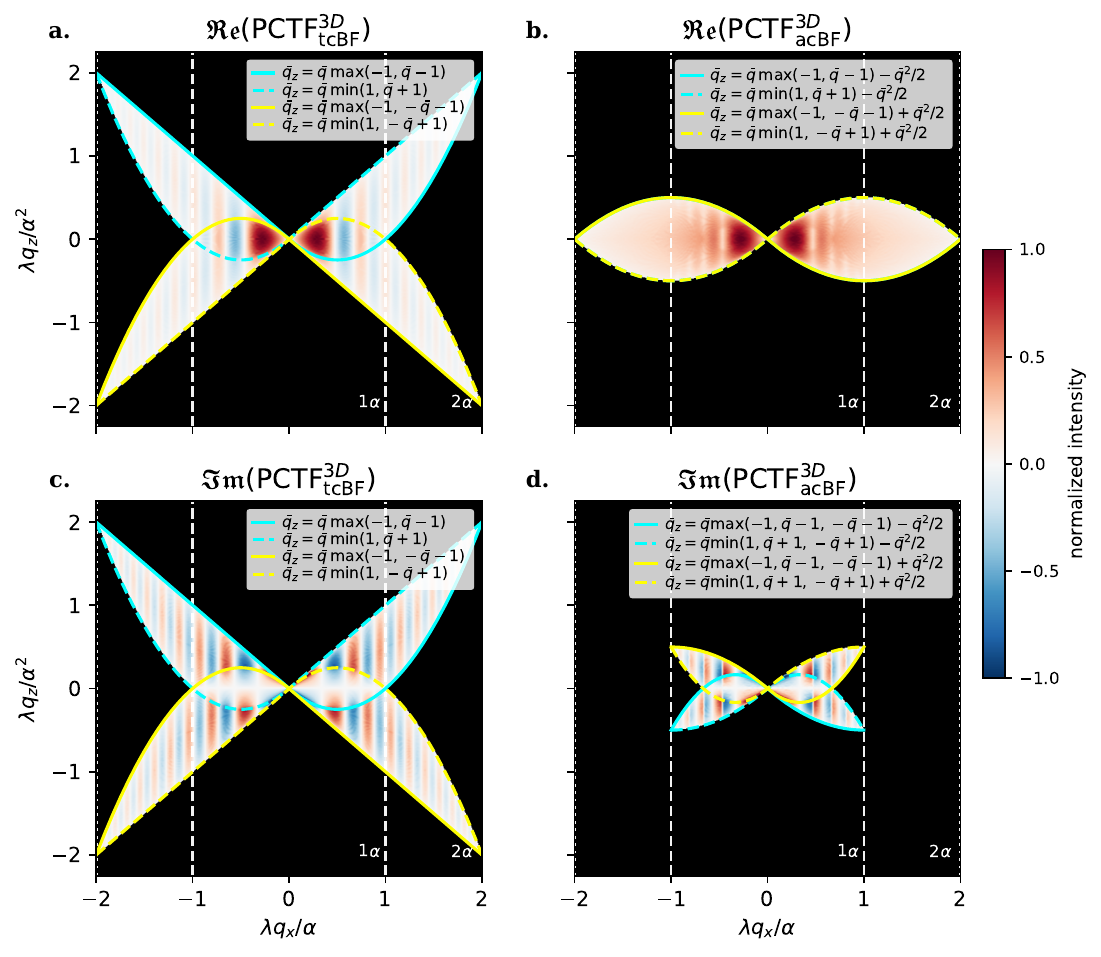}
    \caption{Analytic complex 3D PCTF profiles for tcBF and acBF at 10~nm defocus.
    (a,b) Real parts of the tcBF and acBF PCTFs, respectively. (c,d) Corresponding imaginary parts. The analytic curves show the main envelope boundaries in the $q_y=0$ slice. tcBF follows a broad parallax wedge, while acBF remains on Ewald-sheet support with a narrower DSB region that governs the imaginary, defocus-oscillatory component. Here the barred $\bar{q}$ and $\bar{q}_z$ are spatial frequencies normalized to the aperture size respectively (see details in \ref{appendix_B}).
    }
    \label{fig:tcbf_acbf_3d_ctf_boundaries}
\end{figure}

As above, we show slices of this 3D transfer function numerical FFT of a reconstructed single atom in Figures~\ref{fig:single_atom_analytic_ctf}b and from the theoretical expression in \ref{fig:tcbf_acbf_3d_ctf_boundaries}b,d. 
The suppression of intensity at higher frequencies in the numerical result is due to the finite extent of the carbon atom used in the simulation, whose finite bandwith also modulates the intensity of the 3D Fourier transform. 
The small axial bandwidth near zero lateral frequency accounts for the elongation of the atomic responses seen in Fig.~\ref{fig:two_atom_sectioning}, corresponding to a broad missing wedge of angular information.

The resemblance to an ADF or integrated center-of-mass (iCOM) focal series has a common optical origin. 
In the incoherent ADF approximation, the object's scattering strength is imaged through the probe-intensity point response; in the thin-phase-object limit, ideal iCOM images the specimen phase through the probe intensity instead \citep{xin_2010_ThreeDimensionalImagingAberrationCorrected, lazic_2016_PhasecontrastSTEM}.
The 3D Fourier transform of this probe intensity pairs two plane waves within the illumination aperture. Their lateral frequency is their angular difference divided by $\lambda$, and their axial frequency is the difference of their squared angles divided by $2\lambda$.
These are the same aperture-overlap and Ewald-curvature constraints that give the acBF envelope in Eq.~\eqref{eq:main_acbf_full_support}.
Thus the closed support and depth-localized contrast of acBF resemble focal-series imaging more closely than the broad, shift-only tcBF response.
However, the values of the PCTF within the envelope differ considerably among these methods.
The ADF CTF cannot be directly compared as it has a different dependence on the object.
The DPC PCTF is the same as that for acBF at zero defocus; operating at nonzero defocus improves the acBF PCTF above this baseline. 
Additionally, an ADF or iCOM focal series requires measurements at different physical focus settings, whereas the acBF depth stack is reconstructed from one 4D-STEM acquisition.
This yields substantially higher dose efficiency compared to serially sectioning approaches.
Earlier results on focal-series STEM and SCEM 3D CTFs, and the confocal interpretation of
ptychography, are discussed in Section~\ref{sec:earlier_3d_ctfs}
\citep{xin_2010_ThreeDimensionalImagingAberrationCorrected,
terzoudis-lumsden_2023_ResolutionVirtualDepth}.

\section{Results}\label{sec:results_discussion}
To demonstrate the depth sectioning capabilities of acBF, we performed 3D reconstructions of two experimental datasets obtained from materials with multiple crystalline regions overlapping along the beam direction. 

\subsection[Resolving a twisted bilayer SrTiO3 structure]{Resolving a twisted bilayer SrTiO$_3$ structure}

\begin{figure}[!htbp]
    \centering
    \includegraphics[width=\textwidth]{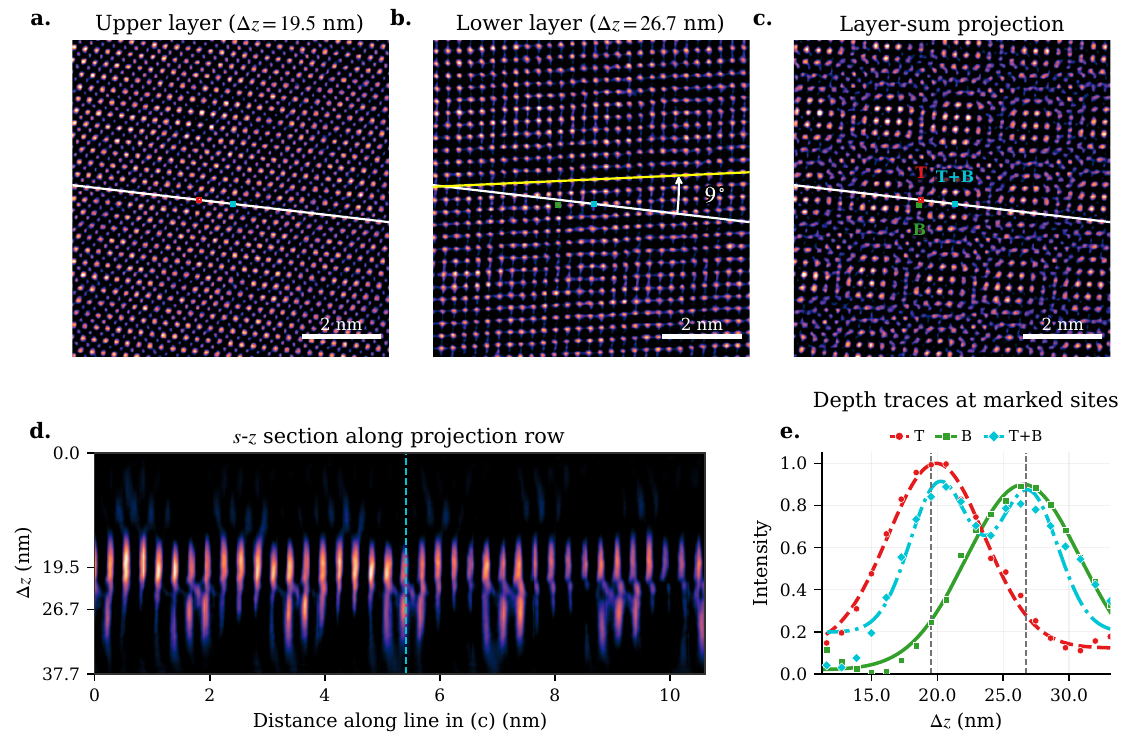}
    \caption{Experimental acBF depth sectioning of a twisted SrTiO$_3$
    bilayer using the data from \citep{kp_2026_Mindthegap}. (a,b) Depth sections at $\Delta z=19.5$ and
    $26.7$~nm. The yellow guide and curved arrow in (b) illustrate the
    estimated $9^\circ$ twist angle relative to the white section line.
    (c) Linear sum of the two sections.
    Red (T) and green (B) boxes
    mark neighboring atomic columns at the boundary between adjacent AA-like
    moir\'e vortices; cyan (T+B) marks an aligned site within one vortex. Each
    box is a $4\times4$-pixel integration region used for the depth trace. (d) Depth
    section displayed from $\Delta z=0$ at the top to $37.7$~nm at the bottom along the white solid line in
    (a--c); the cyan dashed line at $s\approx5.4$~nm passes through T+B. The
    apparent depth extent is roughly 6~nm per layer. (e)
    Normalized depth traces at the marked sites. Circles, squares, and
    diamonds are measured T, B, and T+B samples. Dashed, solid, and dash-dotted
    curves are their Gaussian fits, one for T and B and
    two for T+B. Gray dashed lines mark the displayed sections. The lateral
    sampling is 0.2075~\AA{} on the upsampled grid (0.415~\AA{} scan step),
    and the depth increment is 3.78~\AA{}. Here $\Delta z=0$ is referred to the beam crossover.}
    \label{fig:experimental_sto_depth}
\end{figure}

We applied acBF refocusing to the experimental 4D-STEM dataset from a twisted SrTiO$_3$ bilayer  from KP et al \citep{kp_2026_Mindthegap}. 
The dataset was recorded at 300~kV with a 29.5~mrad convergence semi-angle and 0.415~\AA{} scan sampling. 
The reconstructed volume contains 101 uniformly spaced depth sections from $\Delta z=0$ to $37.7$~nm with a spacing of 3.78~\AA{}.
During reconstruction, each virtual-BF image was upsampled by a factor of 2 to exceed the Nyquist sampling limitation \citep{ma_2026_Information4DSTEMWhere,yu_2022_DoseefficienttcBFSTEMInformation}.
The resulting $512\times512$ grid has 0.2075~\AA{} sampling. 
The two rotated lattice contributions are most clearly separated at $\Delta z=19.5$ and $26.7$~nm. 
Their linear sum provides a projection reference containing both layers and forms a moir\'e pattern because of the $9^\circ$ twist angle (Fig.~\ref{fig:experimental_sto_depth}).
A faint ``ghost'' image of a third layer appears above the sample, but this is likely an artifact from the breakdown of the weak phase approximation used in this method; no such layer appears in the MEP reconstruction. 

Following the top-layer, bottom-layer, and aligned-column comparison used for the same data in multislice electron ptychography~\citep{kp_2026_Mindthegap}, we selected three atomic columns from the acBF stack and perform depth tracing. 
T and B are neighboring atomic columns at the boundary between adjacent AA-like moir\'e vortices, while T+B is a laterally aligned column over the two layers within one vortex. 
Each depth trace integrates the marked $4\times4$-pixel box on the twofold reconstruction grid, equivalent in area to a $2\times2$-pixel box on the native scan grid. 
The T and B traces peak at $\Delta z=19.5$ and $26.7$~nm, respectively, whereas the aligned T+B trace contains both maxima with a minimum between them. 

We performed a 10--90\% knife-edge analysis of the measured T and B depth traces similar to \cite{kp_2026_Mindthegap}. 
Each trace was normalized between its out-of-focus background, taken as the median over $\Delta z=33.1$--$37.7$ and $0.0$--$8.2$~nm, and its measured peak.
Linear interpolation between adjacent depth samples located the threshold crossings. 
The rising-edge 10--90\% and falling-edge 90--10\% widths are 5.79 and 4.21~nm for T, and 5.46 and 5.43~nm for B, respectively.

For comparison, the acBF envelope in Figs.~\ref{fig:single_atom_analytic_ctf} and \ref{fig:tcbf_acbf_3d_ctf_boundaries} contains spatial frequencies up to $|q_z|_{\max}=\alpha^2/(2\lambda)$
(Eq.~\eqref{eq:main_acbf_full_support}).
At 300~kV ($\lambda=1.969$~pm) and $\alpha=29.5$~mrad, the shortest axial period admitted by this envelope is $1/|q_z|_{\max}=2\lambda/\alpha^2=4.52$~nm.
The measured 10--90\% transition widths of 4.21--5.79~nm are consistent with this characteristic scale. These acBF depth resolution measurements are broader than the 2.31 and 2.46~nm knife-edge widths reported for the two layers by KP et al.~\citep{kp_2026_Mindthegap} using multislice electron ptychography, which also uses electrons scattrered to higher angles to achieve greater depth resolution.
A discussion of improved depth resolution from using dark field electrons is given in \ref{appendix_C_confocal_comparison} and \citep{chen_2024_Imaginginterstitialatoms}.
While the present acBF reconstruction does not reveal the same oxygen-sublattice or interfacial-gap detail, acBF recovers the same basic twisted-bilayer arrangement and separates its two lattice contributions, while benefiting from significantly faster speed allowing for live viewing and reconstruction. 

\FloatBarrier
\subsection{Resolving the spatial distribution of Pt nanoparticle grains along depth}

\begin{figure}[!htbp]
    \centering
    \includegraphics[width=\textwidth]{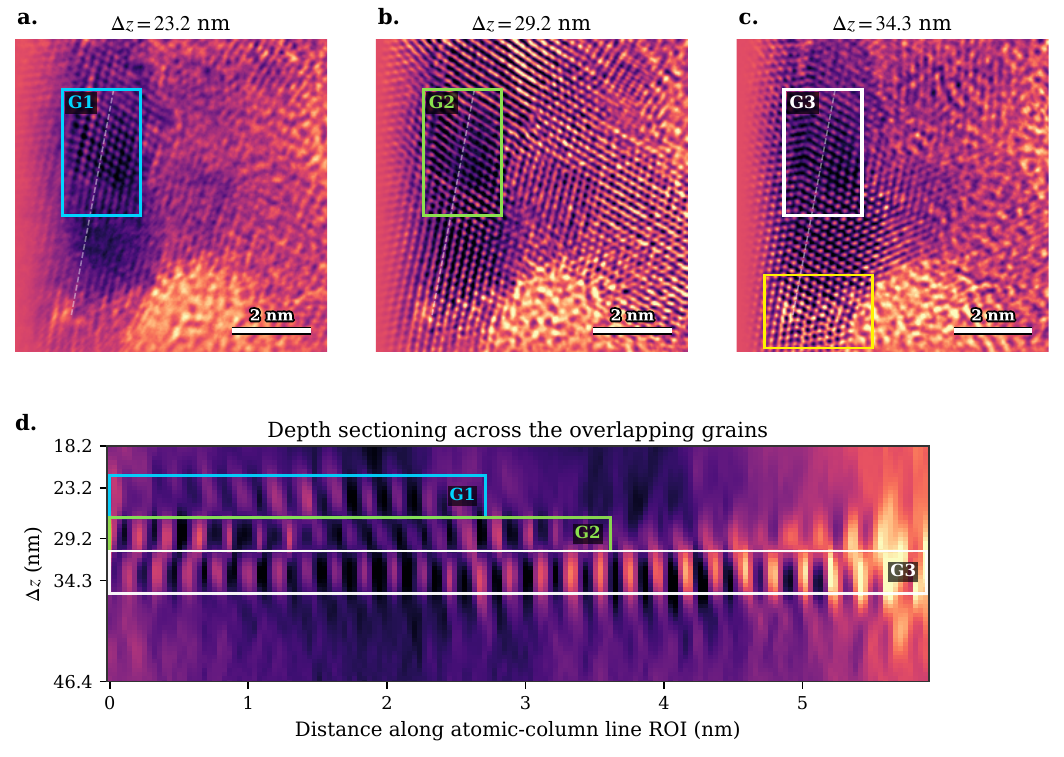}
    \caption{Experimental tcBF depth sectioning of overlapping Pt nanoparticle grains embedded in amorphous
    carbon. (a--c) Depth sections at $\Delta z=23.2$,
    29.2, and 34.3~nm, where
    three vertically overlapping grains G1, G2, and G3 are
    preferentially resolved. (d) Depth sectioning along the dashed line in
    (a--c). Matching boxes mark the G1, G2, and G3 bands over lateral
    distances 0--2.7, 0--3.6, and 0--5.9~nm and depth intervals 21.7--26.7,
    26.7--30.8, and 30.8--35.8~nm, respectively. Here $\Delta z=0$ is referred to the beam crossover. The yellow box indicates an area where the contrast has reversed due to multiple scattering effects in the thicker Pt areas.}
    \label{fig:experimental_pt_depth}
\end{figure}

We next applied tcBF refocusing to a dataset obtained from Pt nanoparticle grains embedded in amorphous carbon. 
The dataset was acquired with a 50~pA probe, 30~mrad convergence semi-angle, 10~nm defocus, and 0.723~\AA{} scan step. 
The reconstructed volume was produced using 120 evenly spaced depth levels with a depth increment of 5.04~\AA{}. 
During reconstruction, each virtual-BF image was again upsampled by a factor of 2 to exceed the Nyquist scan-grid sampling limitation~\citep{ma_2026_Information4DSTEMWhere,yu_2022_DoseefficienttcBFSTEMInformation}, resulting in a $544\times544$ grid with 0.361~\AA{} real space sampling. 

The projected images of three grains overlap in the bottom-left region of the
field of view. 
Their boundaries and lattice orientations are preferentially resolved at $\Delta z=23.2$, 29.2, and 34.3~nm, labeled G1--G3 in order of increasing relative depth (Fig.~\ref{fig:experimental_pt_depth}). 
To follow their contrast continuously through the stack, we placed a fixed, three-pixel-wide line ROI along a nearly vertical atomic-column row visible at $\Delta z=29.2$~nm. 
In the depth section, the G1, G2, and G3 windows extend from the line origin to 2.7, 3.6, and 5.9~nm and span $\Delta z=21.7$--26.7, 26.7--30.8, and 30.8--35.8~nm, respectively. 
The broadened overlap of the lattice fringes reflects the finite axial response and thus limited depth resolution. 
The displayed origin is a reconstruction convention rather than an absolute specimen surface. 
The depth profile in Fig.~\ref{fig:experimental_pt_depth}(d) shows that at least three grains of different  orientation overlap along the beam direction at different $\Delta z$.
The yellow box highlights a region where a contrast reversal occurs. This is due to the breakdown of WPOA when dealing with this thick and heavy-element sample, which causes off-axial images to have altered contrast due to multiple scattering effects.

\section{Discussion}

\FloatBarrier
\subsection{Higher-order aberrations and depth response}
\label{sec:higher_order_depth}
Higher-order aberrations change both the virtual-image shift and its remaining transfer, as we have previously explored for 2D tcBF and acBF imaging \cite{ma_2026_UsingAberrationsImprove}. 
Expanding either sideband phase in Eq.~\eqref{eq:tilt_pctf} to first order in $\boldsymbol q$ gives the common factor $\exp[-i\nabla_{\boldsymbol\Theta}\chi\cdot(\lambda\boldsymbol q)]$.
The corresponding real space image shift is thus given by the gradient of the aberration function \citep{lupini_2016_Rapidaberrationmeasurement,ma_2026_UsingAberrationsImprove}: 
\begin{equation}
\Delta\mathbf r_\chi(\boldsymbol\Theta)
=\frac{1}{k_0}\nabla_{\boldsymbol\Theta}\chi(\boldsymbol\Theta).
\label{eq:general_aberration_shift}
\end{equation}
As discussed earlier, for defocus only this gives $\Delta\mathbf r_\chi=-(\Delta f-z)\boldsymbol\Theta$.
For higher-order aberrations the shift generally varies nonlinearly with detector angle, so a trajectory in the will be curved.
Full expansion of the sideband phases yields terms of every order up to the highest-order aberration coefficient considered, and these terms produce image distortions and substantial changes in contrast transfer for the off-axis detector pixels \cite{ma_2026_UsingAberrationsImprove}.
Example sinograms for coma, three-fold astigmatism and spherical aberration are shown in \ref{appendix_F_sinograms}.

At each depth, we evaluate $\chi_z(\boldsymbol\Theta)=\chi_{\mathrm{ho}}(\boldsymbol\Theta)-k_0(\Delta f-z)|\boldsymbol\Theta|^2/2$, where $\chi_{\mathrm{ho}}$ contains the non-defocus aberrations, and register the virtual images using
\begin{equation}
J_z^{\mathrm{tcBF}}(\mathbf r)
=\int_{|\boldsymbol\Theta|<\alpha}
D\!\left(\mathbf r+\Delta\mathbf r_{\chi_z}(\boldsymbol\Theta),
\boldsymbol\Theta\right)\dd^2\boldsymbol\Theta.
\label{eq:general_aberration_projection}
\end{equation}
Terms beyond first order in $\boldsymbol q$ distorts the image contrast and
are not removed by the rigid shift, leading to greatly suppressed contrast in the final images \citep{ma_2026_UsingAberrationsImprove}. 
acBF corrects their combined phase using the full PCTF, as in Eq.~\eqref{eq:acbf_correction_phasor}, with the shift factor replaced by $\exp[i(2\pi\boldsymbol q)\cdot\Delta\mathbf r_{\chi_z}]$.


In the case of spherical aberration $C_{30}$, off-axial virtual bright field images gain extra defocus together with astigmatism. 
In acBF reconstruction, correcting the PCTF phases before summation amounts to removing the effective defocus changes among the different virtual images, recovering good information transfer even under the influence of aberrations. 
\hyperref[appendix_E_aberrations]{Appendix E} illustrates this distinction for $C_s=2.7$~mm at 300~kV and a 10~mrad convergence semi-angle, at zero defocus and at $\Delta f=89.2$~nm near the Scherzer condition.
The analytical calculations retain the same geometric support boundaries as the corresponding defocus-only expressions, but redistribute the complex transfer within them. 
Such changes can alter axial localization and sidelobes even when the support is unchanged.
Contrast in the tcBF PCTF is almost entirely suppressed except at very low frequencies, with slight improvement when using Scherzer focus.
acBF corrects the residual PCTF phase before summation, reducing phase cancellation between angular
channels and maintaining good contrast transfer throughout the envelope in both cases. 
This is especially evident from the single atom reconstructions shown in \ref{appendix_F_sinograms}, where the atom is completely unrecognizable in the tcBF reconstructions even at infinite dose but is still sharply resolved by acBF. 

\subsection{Relation to 3D CTFs of focal-series STEM and SCEM}
\label{sec:earlier_3d_ctfs}
Three-dimensional contrast transfer functions have previously been derived for several other STEM-based techniques. 
Xin and Muller derived 3D CTFs for ADF-STEM, bright-field scanning confocal electron microscopy (BF-SCEM), and annular dark-field SCEM (ADF-SCEM) \citep{xin_2010_ThreeDimensionalImagingAberrationCorrected}. 
Their ADF-STEM model is an incoherent intensity model in which a three-dimensional scattering-strength distribution is convolved with the probe-intensity PSF; it is therefore not a linear phase contrast CTF. For coherent SCEM, they instead propagate the object through prespecimen and postspecimen lens PSFs, form the intensity at a point detector, and then apply the WPOA. 
The resulting linear phase CTF is the Fourier transform of the interference term between the unscattered reference and the first-order scattered wave, with its envelope constructed from intersections of Ewald surfaces rocked over the illumination and collection apertures. 
This construction established the same finite-angle missing-cone geometry and axial elongation of extended objects that appear in the present analysis.
A similar approach has also been used to derive the 3D CTF of iCoM depth sectioning \cite{bosch_2019_AnalysisDepthSectioningSTEM}, which shares the same envelope shape.

These earlier CTFs describe focal-series STEM or physical SCEM sectioning where a three-dimensional reconstruction is formed by changing the physical focus and recording a series of images. 
The present tcBF/acBF CTFs describe a different measurement and processing order, instead recording the intensity at every diffraction-plane coordinate $\boldsymbol{\Theta}$ separately. 
The SCEM CTF describes coherent angular recombination by a physical postspecimen lens before intensity detection, whereas our CTFs describe computational recombination of detector-resolved intensities. 
The position--angle correlation stores the necessary defocus diversity in one 4D-STEM dataset, so the complete tcBF/acBF depth stack is reconstructed from a single acquisition rather than a focal series.

A later treatment by Terzoudos-Lunsden et al mapped ptychography to an ideal phase-sensitive confocal microscope \citep{terzoudis-lumsden_2023_ResolutionVirtualDepth}.
That correspondence usefully bounds the support that a recovered complex field could reach, but it is not a quantitative ptychographic CTF: the lower postspecimen aperture is treated as uniformly illuminated in the ideal confocal model, whereas ptychography populates and recovers it with nonuniform weights.
This distinction is especially important for dark-field contrast. 
A linear phase term requires a nonzero coherent reference in the intensity expansion $2\mathfrak{Re}[\Psi^{(0)*}\Psi^{(1)}]$. 
A SCEM collection pupil can produce linear phase contrast when its coherently recombined field retains such a reference wave. 
Outside the brightfield disk there is no reference wave, so the cross term vanishes and the leading elastic dark-field intensity is $|\Psi^{(1)}|^2\sim V^2$. 
The same restriction applies directly to a 4D-STEM detector pixel outside the bright-field disk. 
Our derivation instead  starts from the intensity at each detector pixel and retains the aperture products before the tcBF or acBF angular sum. 
\ref{appendix_C_confocal_comparison} compares their reference conditions with the first-order BF intensity response used here.

Iterative ptychography can use dark-field data to recover higher lateral and axial frequencies \citep{chen_2024_Imaginginterstitialatoms, ma_2026_Information4DSTEMWhere}, but that signal has no linear phase CTF and its recovery generally requires an iterative treatment \citep{ma_2026_Information4DSTEMWhere}.
Beyond first order, amplitude contrast can add to or oppose the phase contrast. 
The amplitude contribution discussed in Ref.~\citep{ma_2026_Information4DSTEMWhere} is even in defocus, whereas the uncorrected phase contribution is odd; their combination can enhance contrast at selected frequencies with overfocus and suppress it with underfocus. 
This is not a universal improvement or degradation of the 3D response. 
Higher-order terms become more important for stronger scattering, including heavier atoms and greater thickness, but do not follow a universal atomic-number power law in a coherent crystalline specimen.
Multislice ptychography can account for multiple scattering, as demonstrated in Ref.~\citep{chen_2021_Electronptychographyachieves}; that recovery is distinct from the first-order CTF derived here, and depth resolution beyond the $\lambda/\alpha^2$ first-order limit has been demonstrated experimentally.

\subsection{Limited-angle cone-beam tomography and the missing cone}
\label{sec:limited_cone_tomography}
Optical lightfield reconstruction can also be formulated as synthetic-aperture imaging, wave-optical deconvolution, or limited-angle cone-beam tomography \citep{levoy_2004_Syntheticapertureconfocal,broxton_2013_Waveopticstheory, tian_2015_3Dintensityphase,vigano_2018_Tomographicapproachquantitative, vigano_2019_Advancedlightfieldrefocusing}. 
In the same geometric sense, the detector-resolved virtual images in 4D-STEM provide projection-like views within the angular cone subtended by the probe-forming aperture \citep{seifer_2026_Shadowmontageconebeam}. 

In conventional single-axis tilt tomography, the Fourier planes sampled by the projections rotate about one fixed tilt axis.
For a symmetric tilt range $-\theta_{\max}$ to $+\theta_{\max}$, the unmeasured region is a wedge bounded by planes containing that axis.
Its opening in a section perpendicular to the tilt axis is $180^\circ-2\theta_{\max}$; a typical $\pm70^\circ$ tilt range therefore leaves a $40^\circ$ missing wedge \citep{hovden_2014_BreakingCrowtherLimit}.
Acquiring a second tilt series about an orthogonal axis reduces the remaining missing region to a pyramid \citep{arslan_2006_ReducingMissingWedge}.

From Eq.~\eqref{eq:main_ewald_sheets}, the parallax contribution to the axial spatial frequency obeys $|\boldsymbol{\Theta}\cdot\boldsymbol q| \leq\alpha|\boldsymbol q|$. 
The tcBF envelope therefore forms a wedge whose axial bandwidth collapses as $|\boldsymbol q|\rightarrow0$.
The missing information around the $q_z$ axis, which in a central $q_x$--$q_z$ section appears as a  missing wedge, is in fact a missing cone due to the rotational symmetry of the 3D CTF.
Thus, while we have effectively sampled a full distribution of tilt \emph{axes}, the total tilt range on any tilt axis is only $\pm\alpha$.
For $\alpha=30$~mrad ($1.7^\circ$), the transferred band spans only about $3.4^\circ$ in a central section, and the missing cone has an opening of about $176.6^\circ$ \citep{intaraprasonk_2008_Analyticderivationoptimal, hovden_2014_BreakingCrowtherLimit}.
The missing information is responsible for the anisotropic resolution, axial elongation, and sidelobes, especially for laterally extended or slowly varying objects \citep{behan_2009_Threedimensionalimagingoptical, xin_2010_ThreeDimensionalImagingAberrationCorrected}.

Increasing $\alpha$ closes the cone and increases the available axial bandwidth, but it does not make a single-view reconstruction isotropic.
While iterative, non-linear methods can utilize the dark-field scattered electrons, simply extending the collection angle beyond the bright-field disk also does not enlarge the linear phase CTF, as shown in \ref{appendix_C_confocal_comparison}. 
The single-acquisition depth stacks are therefore best interpreted as anisotropic, PCTF-limited reconstructions rather than complete tomograms.
If a tomographic series of 4D-STEM datasets is acquired, the depth information present in each scan can be utilized to more effectively fill reciprocal space information.
This has been demonstrated with multislice ptychography by fusing multiple MEP acquisitions at different sample tilts via simultaneous reconstruction \citep{dong_2025_TiltCoupledMultislicePtychography, you_2025_EndToEndPtychographicTomography}. 
Exploiting the depth information within each scan could reduce the number of tilted acquisitions required for a given resolution, potentially allowing more efficient use of the limited dose budget for frozen biological samples \citep{yu_2024_Exploring4DSTEMCryoET,yu_2025_AdvancingCryoETtcBF}.

\section{Conclusion}\label{sec:conclusion}

Depth information is encoded in aberrated 4D-STEM acquisitions as parallax among detector-resolved virtual images. 
Defocus produces a shift linear in detector angle, but linearity is not required: in general the shift follows the gradient of the aberration phase. 
Applying the inverse angle-dependent shift field for a depth registers one axial plane at a time. 
Repeating this shifting process for different axial planes produces a depth section without the need for a focal or tilt series, allowing for straightforward 3D reconstruction with tcBF and acBF.
tcBF reconstruction corrects the parallax shift but not the contrast transfer changes, while acBF also aligns the complex, tilt-dependent phase transfer.

In this work, we have developed analogies between 4D-STEM 3D imaging and two established 3D imaging methods.
By analogy with tomography, we show how the sinogram construction provides a straightfoward visualization of parallax information encoding that can be applied to 4D-STEM datasets.
The conventional tcBF reconstruction algorithm is equivalent to unfiltered backprojection tomography algorithm under the small-angle approximation.  
By analogy with lightfield imaging, we show that 4D-STEM is an efficient and scalable way of acquiring position--angle information via its serial diffraction pattern acquisition.
Adapting the 4D-FFT approach from lightfield reconstruction, we have shown that substantial speedup of 3D volume reconstruction from tcBF is possible when reconstructing many depth planes. 
However, the wave-optical nature of phase contrast imaging deviates from the typical assumptions in LF imaging, so that this algorithm is not directly applicable to acBF reconstruction. 
Further development of accelerated algorithms may eventually enable live volume reconstruction with these direct ptychography approaches. 

The limited probe-forming aperture in 4D-STEM leaves a large missing low-frequency region in the 3D CTF and thus an elongated axial response in the point spread function. 
Scan and detector sampling, noise, partial coherence, multiple scattering, and residual aberrations further limit quantitative recovery outside the weak-phase regime. 
Within these limits, parallax refocusing and the 3D PCTF provide complementary geometric and wave-optical descriptions of three-dimensional information recovered from one 4D-STEM measurement.

The non-iterative 3D reconstructions shown here are promising candidates for applications where samples are weakly scattering and in the low-dose regime where the nonlinear information that enables superresolution is lost to measurement noise. 
This is particularly the case in cryo electron tomography, where we envision that 3D reconstructions will enable the use of fewer tilt angles by filling in a wedge of 3D information from each scan rather than simply combining projection images. 
For samples much thicker than the depth of field, these 3D reconstruction methods also enable extended depth of field reconstruction that being the entire sample into focus at every projection. 

\section*{CRediT authorship contribution statement}
\textbf{Desheng Ma:} Conceptualization, Methodology, Software, Formal analysis,
Investigation, Data curation, Visualization, Writing -- original draft,
Writing -- review \& editing. 
\textbf{Chia-Hao Lee:} Methodology, Software,
Writing -- review \& editing. 
\textbf{Zixiao Shi:} Data collection, Investigation,
Writing -- review \& editing. 
\textbf{David A. Muller:} Conceptualization, Supervision, Funding
acquisition, Writing -- review \& editing. 
\textbf{Steven E. Zeltmann:} Conceptualization,
Methodology, Software, Supervision, Writing -- original draft, Writing -- review \&
editing.

\section*{Data and code availability}
The public acBF/tcBF reference implementation is available at \url{https://github.com/dsmagiya/acBF-STEM}.
A GPU-accelerated version of acBF is available at \url{https://github.com/chiahao3/fast-acbf}; this version can be run interactive in the py4DSTEM GUI via the plugin available at \url{https://github.com/chiahao3/py4d-browser-fast-acbf}.
The simulation inputs and processed experimental depth stacks supporting this study are available from the corresponding authors upon reasonable request.

\section*{Declaration of competing interest}
The authors declare that they have no known competing financial interests or personal relationships that could have appeared to influence the work reported in this paper.

\section*{Acknowledgments}
We thank Laura Waller for helpful discussions on light-field imaging.  
DM was supported by the Center for Bright Beams, a U.S. National Science Foundation Science and Technology Center under Award PHY-1549132. 
SEZ is supported by Army Research Office grant W911NF-25-1-0256.
Electron microscopy facility support was provided by the Cornell Center for Materials Research Shared Facilities with support through the NSF Materials Research Science and Engineering Centers program under Award DMR-1719875.


\appendix
\FloatBarrier
\input{appendix_lf_tcbf_benchmark}
\input{appendix_rose_3d_pctf}
\input{appendix_support_boundaries}
\FloatBarrier
\input{appendix_confocal_comparison}
\input{appendix_spherical_aberration}
\input{appendix_aberration_sinograms}

\section*{Declaration of generative AI use}
During the preparation of this work, the authors used OpenAI Codex to assist with language editing, manuscript organization, consistency checks, and plotting-code preparation. 
The authors reviewed and edited the resulting content as needed and take full responsibility for the content of the publication.

\begingroup
\sloppy
\hbadness=10000
\bibliographystyle{elsarticle-num}
\bibliography{lightfield}
\endgroup

\end{document}

%% file: appendix_lf_tcbf_benchmark.tex
\renewcommand{\thefigure}{A\arabic{figure}}
\renewcommand{\theHfigure}{A.\arabic{figure}}
\setcounter{figure}{0}
\renewcommand{\theequation}{A\arabic{equation}}
\renewcommand{\theHequation}{A.\arabic{equation}}
\setcounter{equation}{0}
\renewcommand{\thetable}{A\arabic{table}}
\renewcommand{\theHtable}{A.\arabic{table}}
\setcounter{table}{0}
\section{Computational cost of LF slicing and tcBF}
\label{appendix_D_benchmark}

\begin{figure}[!htbp]
\centering
\includegraphics[width=\textwidth]{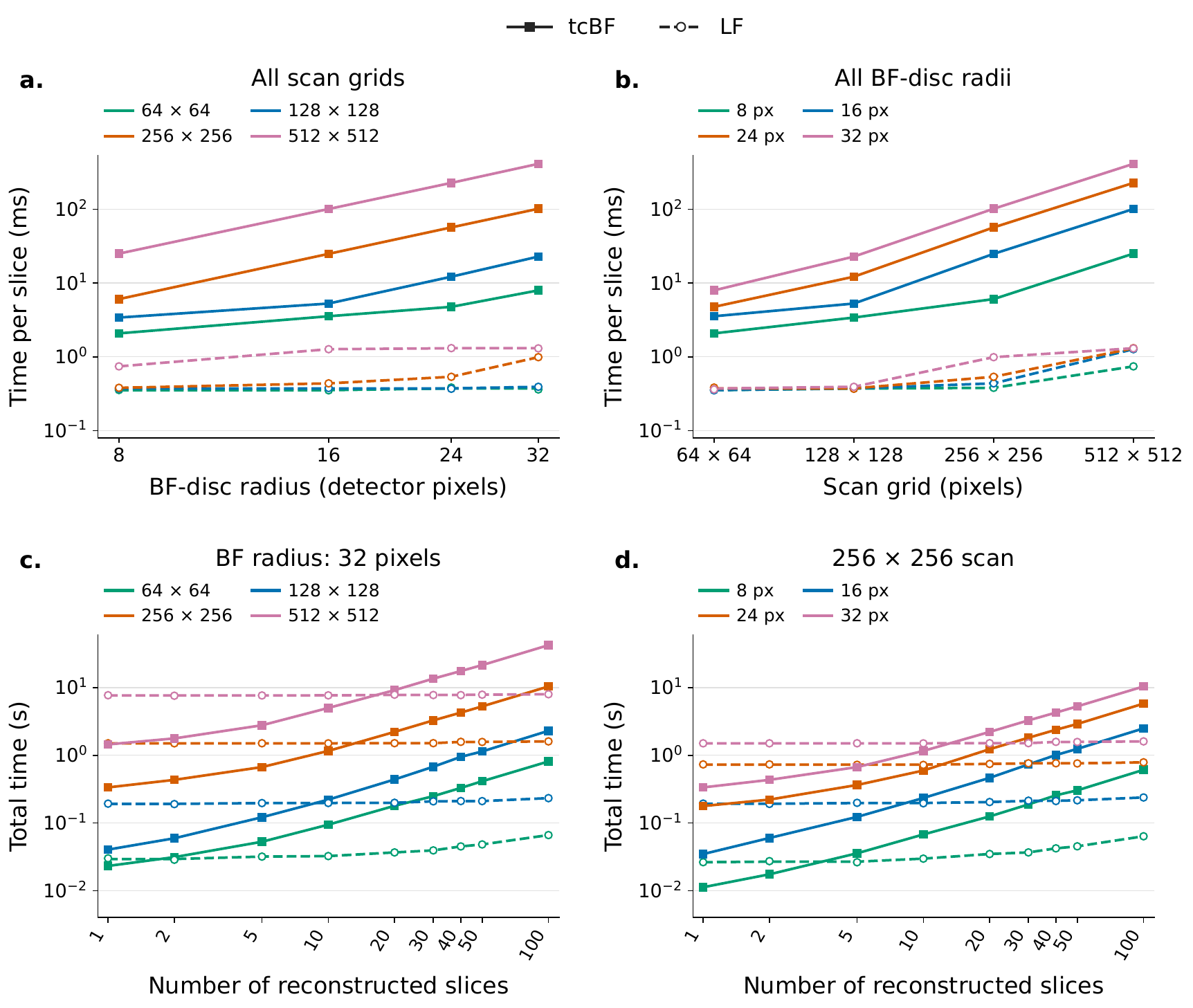}
\caption{LF and tcBF computation time.
(a,b) Reconstruction time per slice versus BF-disc radius and
scan size, respectively.
(c,d) Total time, including 4D-FFT pre-computation, versus number of depth slices at
a 32-pixel BF-disc radius and a $256\times256$ scan, respectively.
Axes are logarithmic. Filled squares and solid lines denote tcBF;
open circles and dashed lines denote LF.
Colors identify scan sizes in (a,c) and BF-disc radii in (b,d).
Symbols show measured means over 50 trials. 
In LF, frequency grids for slicing are cached once per stack during the reconstructions.
}
\label{fig:lf_tcbf_benchmark}
\end{figure}

Figure~\ref{fig:lf_tcbf_benchmark} benchmarks the reconstruction time per slice
after 4D-FFT precomputation and the total time needed to produce a depth stack.
The upper row covers all combinations of four scan grids
($64\times64$, $128\times128$, $256\times256$, and $512\times512$)
and four BF-disc radii (8, 16, 24, and 32 detector pixels).
Panel (a) varies BF-disc radius, with a separate series for each scan size;
panel (b) varies scan size, with a separate series for each BF-disc radius.
The lower row varies the number of reconstructed slices, at a fixed BF-disc
radius of 32 pixels in (c) and a fixed $256\times256$ scan in (d).
The radii correspond to 195, 795, 1791, and 3207 pixels within a circular
30~mrad aperture on $16\times16$, $32\times32$, $48\times48$, and
$64\times64$ detector grids, respectively. Increasing the radius in pixels
improves angular sampling while the convergence semiangle remains the same (30 mrad in the simulation).
All cases use a 0.5~\AA{} scan step,
a 0.019687~\AA{} wavelength, and no lateral upsampling.
Only effective defocus is varied; all other aberration coefficients are zero.
Measurements use an Apple M2 Max GPU with 64~GiB unified memory and
PyTorch~2.11.0 (Metal backend).

Each method runs in a separate process, with alternating method order.
The configuration order is randomized independently in each session.
After one untimed initialization run, Fourier and geometry caches and unused GPU memory
are cleared before every trial.
The full protocol comprises five sessions of ten repeats, giving 50 measurements per point.
Slice counts are tested in randomized order, with GPU synchronization at
the start, end, and setup-stage boundaries.
The upper row uses the measured reconstruction stage of the 100-slice trials,
divided by 100, excluding any pre-computation time.

Timing starts with virtual BF images on the GPU.
LF setup includes native-grid embedding, the centered 4D FFT, and construction
of the spatial-frequency grids and angular-frequency spacings.
These depth-independent quantities are computed once per stack and reused
for every slice; only the depth-dependent sampling coordinates are updated.
For the native 4D FFT, frequency shifts are applied one axis at a time to
release temporary buffers between shifts.
Reconstruction includes Fourier slicing and inverse 2D FFTs for LF, or
shift-and-sum processing in batches of 64 BF pixels for tcBF.
Output stacking is included; common input extraction, initial device transfer,
and disk I/O are excluded.
Every scan size uses the same native 4D FFT for LF and the same native
spatial-FFT cache for tcBF; there is no size-specific
FFT batching. GPU allocation is
limited to 40~GiB, with available-memory checks included in the measured
intervals. 

%% file: appendix_rose_3d_pctf.tex
\FloatBarrier
\renewcommand{\thefigure}{B\arabic{figure}}
\renewcommand{\theHfigure}{B.\arabic{figure}}
\setcounter{figure}{0}
\renewcommand{\theequation}{B\arabic{equation}}
\renewcommand{\theHequation}{B.\arabic{equation}}
\setcounter{equation}{0}
\renewcommand{\thetable}{B\arabic{table}}
\renewcommand{\theHtable}{B.\arabic{table}}
\setcounter{table}{0}
\section{3D phase contrast transfer from Rose's formalism}
\label{appendix_A}

\label{appendix_A_rose_pctf}

This appendix accompanies Section~\ref{sec:3d_ctf}.
We retain the Ewald depth terms throughout the 3D treatment. The projected
2D PCTFs are recovered by integrating the complex 3D transfer over $q_z$;
the 2D PCTFs cannot be used as the starting point for a depth-axis Fourier transform.

We start from the coherent phase term in Rose's generalized contrast
formalism. For a detector pixel at scattering angle
$\boldsymbol{\Theta}$, the coherent phase contrast of the object at depth $z$ may be written as
\begin{equation}
\resizebox{\linewidth}{!}{$
\begin{aligned}
C_{\text{phase}}(\boldsymbol{\rho},z;\boldsymbol\Theta)
={}&
\frac{2}{\lambda\Omega_0}
A(\boldsymbol\Theta)
\mathfrak{Im}
\int
A(\boldsymbol{\theta})
\exp\left\{
i\left[
\chi(\boldsymbol\Theta)
-
\chi(\boldsymbol{\theta})
+
k_0(\boldsymbol{\theta}-\boldsymbol\Theta)\cdot\boldsymbol{\rho}
-
\frac{k_0 z}{2}
\left(
|\boldsymbol{\theta}|^2
-
|\boldsymbol\Theta|^2
\right)
\right]
\right\}
F_s(\boldsymbol{\theta},\boldsymbol\Theta)
\dd^2\boldsymbol{\theta} .
\end{aligned}
$}
\label{eq:rose_coherent_phase}
\end{equation}
Here $\boldsymbol{\theta}$ is the incident angle inside the probe-forming aperture,
$\boldsymbol{\Theta}$ is the detector angle, $A$ is the aperture function,
$\chi$ is the aberration phase, $\boldsymbol{\rho}$ is the transverse
real-space coordinate, $k_0=2\pi/\lambda$,
$\Omega_0$ is the probe-aperture area, and $F_s$ is the symmetric component of the scattering amplitude in Rose's notation \citep{rose_1977_Nonstandardimagingmethods,ma_2026_Information4DSTEMWhere}. The effective defocus is
$\Delta f_{\mathrm{eff}}=\Delta f-z=-\Delta z$.
The term proportional to $z$ is the
Ewald-sphere term, absent in the projected 2D PCTF under paraxial approximation.

Define the transverse spatial frequency (in cycles per unit length) by
\begin{equation}
\boldsymbol q=\frac{\boldsymbol{\theta}-\boldsymbol\Theta}{\lambda},
\qquad
\boldsymbol{\theta}
=
\boldsymbol\Theta+\lambda\boldsymbol q.
\end{equation}
Substitution into the depth term of Eq.~\eqref{eq:rose_coherent_phase}
gives the axial factor for the
$A(\boldsymbol\Theta+\lambda\boldsymbol q)$ sideband:
\[
\exp\!\left[-i2\pi z\left(\boldsymbol\Theta\cdot\boldsymbol q
+\frac{\lambda}{2}|\boldsymbol q|^2\right)\right].
\]
Reindexing its complex conjugate to the same $\boldsymbol q$ gives the
$A(\boldsymbol\Theta-\lambda\boldsymbol q)$ sideband, with the opposite
sign of the quadratic term. Their axial frequencies are therefore
\begin{equation}
{
q_z
=
Q_\pm(\boldsymbol\Theta,\boldsymbol q)
=
\boldsymbol\Theta\cdot\boldsymbol q
\pm
\frac{\lambda}{2}|\boldsymbol q|^2 .
}
\label{eq:ewald_sheets}
\end{equation}

Here $q_z$ is conjugate to object depth $z$ and is also in cycles per unit
length. With the depth dependence written as $e^{-i2\pi q_z z}$, each
sideband factor has the representation
\begin{equation}
\int_{-\infty}^{\infty}
\delta(q_z-Q_\pm)e^{-i2\pi q_z z}\dd q_z
=e^{-i2\pi Q_\pm z}.
\label{eq:axial_delta}
\end{equation}

Factoring out the object spectrum in Eq.~\eqref{eq:rose_coherent_phase}
and combining the two sidebands gives the detector-pixel 3D PCTF:
\begin{equation}
{
\begin{aligned}
\PCTF_{\mathrm{3D}}
(\boldsymbol q,q_z;\boldsymbol\Theta)
=
\frac{i}{2\Omega_0}
A(\boldsymbol\Theta)
\Big[
&
A(\boldsymbol\Theta-\lambda\boldsymbol q)
e^{-i[\chi(\boldsymbol\Theta)-\chi(\boldsymbol\Theta-\lambda\boldsymbol q)]}
\delta(q_z-Q_-)
\\
&
-
A(\boldsymbol\Theta+\lambda\boldsymbol q)
e^{+i[\chi(\boldsymbol\Theta)-\chi(\boldsymbol\Theta+\lambda\boldsymbol q)]}
\delta(q_z-Q_+)
\Big].
\end{aligned}
}
\label{eq:3d_pctf_general}
\end{equation}
The aperture products select which (or both) sidebands are present; the aberration
phases modulate the transfer without changing their Ewald-sheet envelopes. This is
the transfer at one detector pixel, before tcBF or acBF correction and
angular summation.

\subsection{Defocus phase and the projected 2D limit}

For defocus only, $\chi(\boldsymbol\theta)=-k_0\Delta f
|\boldsymbol\theta|^2/2$, so the phase difference terms are
\begin{equation}
\chi(\boldsymbol\Theta)
-
\chi(\boldsymbol\Theta\pm\lambda\boldsymbol q)
=
\pm2\pi\Delta f\,\boldsymbol\Theta\cdot\boldsymbol q
+
\pi\lambda\Delta f|\boldsymbol q|^2 ,
\label{eq:defocus_phase_pm}
\end{equation}
Substitution into Eq.~\eqref{eq:3d_pctf_general} gives
\begin{equation}
\resizebox{\linewidth}{!}{$
{
\begin{aligned}
\PCTF_{\mathrm{3D}}^{\Delta f}
(\boldsymbol q,q_z;\boldsymbol\Theta)
=
\frac{i}{2\Omega_0}
A(\boldsymbol\Theta)
e^{+i2\pi\Delta f\,\boldsymbol\Theta\cdot\boldsymbol q}
\Big[
&
A(\boldsymbol\Theta-\lambda\boldsymbol q)
e^{-i\pi\lambda\Delta f|\boldsymbol q|^2}
\delta(q_z-Q_-)
\\
&
-
A(\boldsymbol\Theta+\lambda\boldsymbol q)
e^{+i\pi\lambda\Delta f|\boldsymbol q|^2}
\delta(q_z-Q_+)
\Big].
\end{aligned}
}
$}
\label{eq:3d_pctf_defocus}
\end{equation}
The common factor $e^{i2\pi\Delta f\,\boldsymbol\Theta\cdot\boldsymbol q}$
produces an image shift; $e^{\mp i\pi\lambda\Delta f|\boldsymbol q|^2}$
are the residual sideband factors.
Integrating Eq.~\eqref{eq:3d_pctf_defocus} over $q_z$ gives the
projected 2D PCTF, while retaining the aberration phase differences:
\begin{equation}
\resizebox{\linewidth}{!}{$
{
\begin{aligned}
\PCTF_{\mathrm{2D}}^{\Delta f}
(\boldsymbol q;\boldsymbol\Theta)
={}&\int\PCTF_{\mathrm{3D}}^{\Delta f}
(\boldsymbol q,q_z;\boldsymbol\Theta)\dd q_z\\
={}&
\frac{i}{2\Omega_0}
A(\boldsymbol\Theta)
e^{+i2\pi\Delta f\,\boldsymbol\Theta\cdot\boldsymbol q}
\left[
A(\boldsymbol\Theta-\lambda\boldsymbol q)
e^{-i\pi\lambda\Delta f|\boldsymbol q|^2}
-
A(\boldsymbol\Theta+\lambda\boldsymbol q)
e^{+i\pi\lambda\Delta f|\boldsymbol q|^2}
\right].
\end{aligned}
}
$}
\label{eq:flat_ewald_2d_pctf_defocus}
\end{equation}
%

\subsection{tcBF 3D PCTF}

As in Section~\ref{sec:analytic_tcbf_transfer}, we first return to the
3D expression for the detector-pixel resolved CTF and then apply the depth-dependent shift before
summing over detector angles.
For defocus alone, $\chi(\boldsymbol\Theta)
=-\pi\Delta f|\boldsymbol\Theta|^2/\lambda$, so the two remaining
sideband phase factors are $\exp(i2\pi\Delta fQ_\pm)$.
The tcBF shifting multiplies both terms by
$\exp[-i2\pi(\Delta f-z)\boldsymbol\Theta\cdot\boldsymbol q]$,
where $z$ is the reconstruction depth.
Substituting $Q_\pm$ from Eq.~\eqref{eq:ewald_sheets} and Fourier
transforming the common factor along reconstruction depth gives
\begin{equation}
\begin{aligned}
e^{-i2\pi(\Delta f-z)\boldsymbol\Theta\cdot\boldsymbol q}
e^{i2\pi\Delta fQ_\pm}
&=e^{i2\pi z\boldsymbol\Theta\cdot\boldsymbol q}
e^{\pm i\pi\lambda\Delta f|\boldsymbol q|^2},\\
\int_{-\infty}^{\infty}
e^{i2\pi z\boldsymbol\Theta\cdot\boldsymbol q}
e^{-i2\pi q_z z}\dd z
&=\delta(q_z-\boldsymbol\Theta\cdot\boldsymbol q).
\end{aligned}
\label{eq:tcbf_point_response_mapping}
\end{equation}
Thus both sidebands have the same reconstruction-depth dependence,
while their opposite quadratic Fresnel phases remain.
Including the aperture factors and relative sign of the two sidebands
in Eq.~\eqref{eq:3d_pctf_general}, and summing over detector angles, gives
\begin{equation}
{
\begin{aligned}
\PCTF_{\mathrm{tcBF}}^{\mathrm{3D}}(\boldsymbol q,q_z)
=
\frac{i}{2\Omega_0}
\int \dd^2\boldsymbol{\Theta}\,
A(\boldsymbol{\Theta})
\delta\left(
q_z-\boldsymbol{\Theta}\cdot\boldsymbol q
\right)
\Bigg[
&
A(\boldsymbol{\Theta}-\lambda\boldsymbol q)
e^{-i\pi\lambda\Delta f|\boldsymbol q|^2}
\\
-
&
A(\boldsymbol{\Theta}+\lambda\boldsymbol q)
e^{+i\pi\lambda\Delta f|\boldsymbol q|^2}
\Bigg].
\end{aligned}
}
\label{eq:tcbf_3d}
\end{equation}
Both terms lie on the same Ewald sheet
$q_z=\boldsymbol\Theta\cdot\boldsymbol q$, but retain their opposite
quadratic phases.

Collapsing over $q_z$ gives the usual two-dimensional tcBF PCTF,
\begin{equation}
{
\int
\PCTF_{\mathrm{tcBF}}^{\mathrm{3D}}(\boldsymbol q,q_z)
\dd q_z
=
\mathbb A(\boldsymbol q)
\sin\left(
\pi\lambda\Delta f|\boldsymbol q|^2
\right).
}
\label{eq:tcbf_2d_collapse}
\end{equation}
Here $\mathbb A(\boldsymbol q)$ is the normalized overlap area of two angular aperture disks separated by $\lambda|\boldsymbol q|$.

\subsection{acBF 3D PCTF}

For acBF, the phase correction is determined from the full per-tilt PCTF, not from the two Ewald branches separately.
The expressions below are evaluated at $z=0$, as in
Section~\ref{sec:analytic_acbf_transfer}. At another selected depth, the
correction is evaluated using $\Delta f-z$.
After removing the same common defocus-induced shift phase, the residual
per-tilt PCTF is
\begin{equation}
\begin{aligned}
\PCTF^{\mathrm{res}}_0(\boldsymbol q,\boldsymbol\Theta)
={}&\frac{iA(\boldsymbol\Theta)}{2\Omega_0}\Big[
A(\boldsymbol\Theta-\lambda\boldsymbol q)
e^{-i\pi\lambda\Delta f|\boldsymbol q|^2}\\
&\qquad{}-
A(\boldsymbol\Theta+\lambda\boldsymbol q)
e^{+i\pi\lambda\Delta f|\boldsymbol q|^2}\Big].
\end{aligned}
\label{eq:acbf_per_tilt}
\end{equation}
Thus the acBF phase correction is
\begin{equation}
\resizebox{\linewidth}{!}{$
\mathcal{C}_{\mathrm{acBF}}
(\boldsymbol\Theta,\boldsymbol q)
=
\exp\left\{
-i\arg\left[
i
\left(
A(\boldsymbol\Theta-\lambda\boldsymbol q)
e^{-i\pi\lambda\Delta f|\boldsymbol q|^2}
-
A(\boldsymbol\Theta+\lambda\boldsymbol q)
e^{+i\pi\lambda\Delta f|\boldsymbol q|^2}
\right)
\right]
\right\}.
$}
\label{eq:acbf_phase_factor}
\end{equation}
At zero residual transfer the correction is set to zero, as in
Eq.~\eqref{eq:acbf_correction_phasor}.
The same phase correction must be applied to both branches before the branches are placed on their respective Ewald sheets. Therefore,
\begin{equation}
\resizebox{\linewidth}{!}{$
{
\begin{aligned}
\PCTF_{\mathrm{acBF}}^{\mathrm{3D}}(\boldsymbol q,q_z)
=
\frac{i}{2\Omega_0}
\int \dd^2\boldsymbol\Theta\,
A(\boldsymbol\Theta)
\mathcal{C}_{\mathrm{acBF}}
(\boldsymbol\Theta,\boldsymbol q)
\Bigg[
&
A(\boldsymbol\Theta-\lambda\boldsymbol q)
e^{-i\pi\lambda\Delta f|\boldsymbol q|^2}
\delta\left(
q_z
-
\boldsymbol\Theta\cdot\boldsymbol q
+
\frac{\lambda}{2}|\boldsymbol q|^2
\right)
\\
-
&
A(\boldsymbol\Theta+\lambda\boldsymbol q)
e^{+i\pi\lambda\Delta f|\boldsymbol q|^2}
\delta\left(
q_z
-
\boldsymbol\Theta\cdot\boldsymbol q
-
\frac{\lambda}{2}|\boldsymbol q|^2
\right)
\Bigg].
\end{aligned}
}
$}
\label{eq:acbf_3d}
\end{equation}
The delta functions place the $A(\boldsymbol\Theta\pm\lambda\boldsymbol q)$
branches on $q_z=Q_\pm$ [Eq.~\eqref{eq:ewald_sheets}].
The correction is determined from their combined per-tilt PCTF, not by
taking the magnitude of each branch separately.
The depth-dependent correction is evaluated at $\Delta f-z$ before the
detector sum. The Ewald terms are retained in the 3D expression; no
Fourier transform of the projected 2D depth sweep is used here.

\subsection{SSB and DSB interpretation}

In a single-sideband (SSB) region, only one of
$A(\boldsymbol\Theta\pm\lambda\boldsymbol q)$ is nonzero, so the
per-tilt magnitude is independent of defocus.
In the double-sideband (DSB) region both equal one, giving
\begin{equation}
\left|
e^{-i\pi\lambda\Delta f|\boldsymbol q|^2}
-
e^{+i\pi\lambda\Delta f|\boldsymbol q|^2}
\right|
=
2
\left|
\sin\left(
\pi\lambda\Delta f|\boldsymbol q|^2
\right)
\right|.
\end{equation}
Thus the defocus oscillation in acBF is restricted to the DSB region.

Collapsing Eq.~\eqref{eq:acbf_3d} over $q_z$ gives
\begin{equation}
{
\int \PCTF_{\mathrm{acBF}}^{\mathrm{3D}}(\boldsymbol q,q_z)\dd q_z
=
[
\mathbb A(\boldsymbol q)
-
\mathbb A(2\boldsymbol q)
]
+
\mathbb A(2\boldsymbol q)
\left|
\sin\left(
\pi\lambda\Delta f|\boldsymbol q|^2
\right)
\right|.
}
\label{eq:acbf_2d_collapse}
\end{equation}
Since $\mathbb A(2\boldsymbol q)=0$ beyond $\alpha/\lambda$, the
oscillatory term is confined to $|\boldsymbol q|<\alpha/\lambda$;
for $\alpha/\lambda<|\boldsymbol q|<2\alpha/\lambda$, acBF is purely SSB
and non-oscillatory.

\subsection{Closed-form evaluation for a circular aperture}
\label{sec:closed_form_3d_pctf}

\begin{figure}[!htbp]
    \centering
    \includegraphics[width=0.95\textwidth]{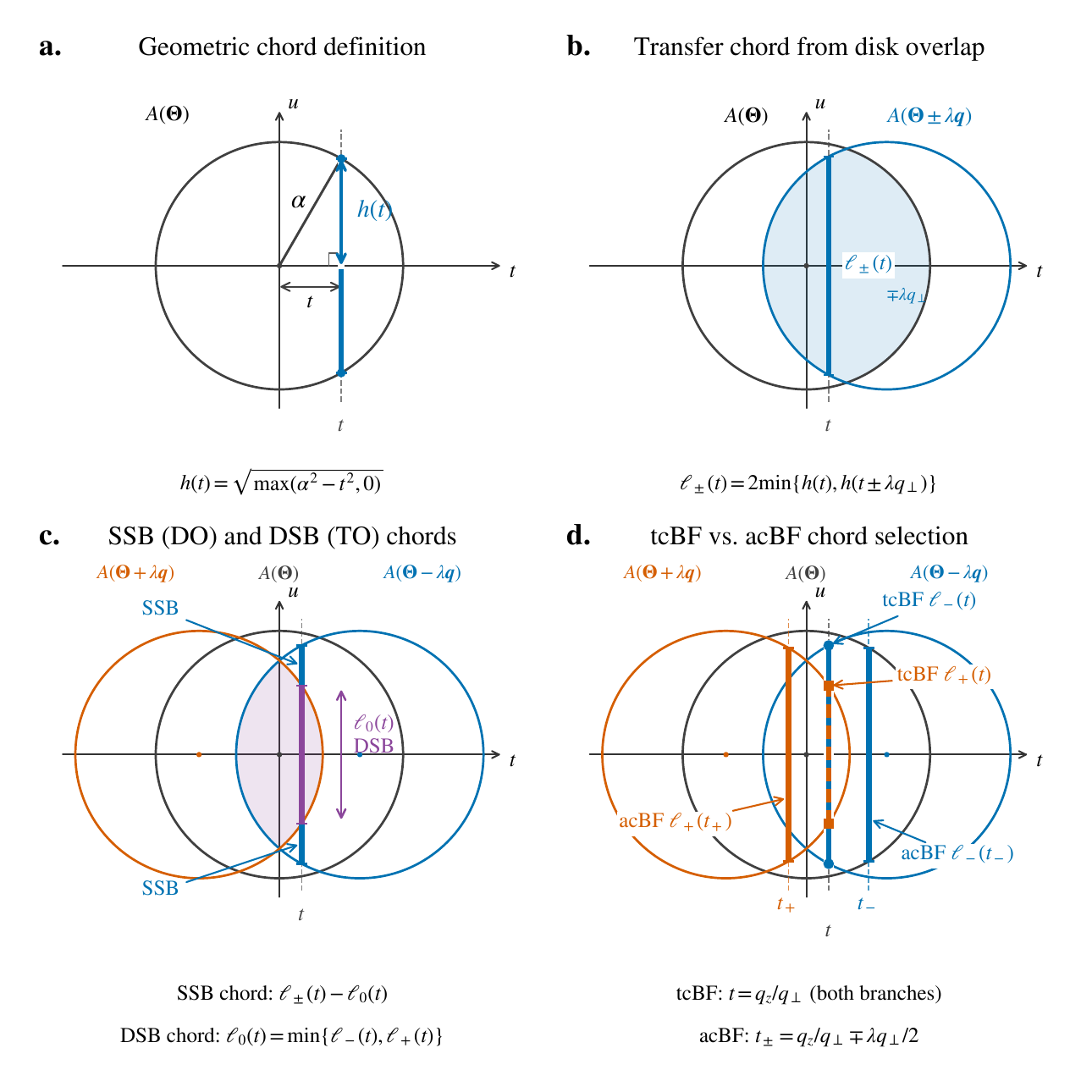}
    \caption{Aperture-overlap chords in the detector-angle plane defined in Fig.~\ref{eq:appendix_aperture_chord_lengths}, with
    $t\parallel\boldsymbol q$ and $u\perp\boldsymbol q$.
    (a) A circular aperture of radius $\alpha$ has half-chord length $h(t)$.
    (b) The overlap with a disk centered at $\mp\lambda q_\perp$ gives
    $\ell_\pm(t)$; the right-shifted disk shown corresponds to $\ell_-(t)$.
    (c) The common triple-overlap (TO) chord $\ell_0(t)$ is the DSB
    contribution; the remaining double-overlap (DO) segments have total
    length $\ell_\pm(t)-\ell_0(t)$ for each sideband.
    (d) tcBF evaluates both chords at the same position
    $t=q_z/q_\perp$, whereas acBF evaluates them at
    $t_\pm=q_z/q_\perp\mp\lambda q_\perp/2$.
    The central solid blue and dashed orange segments mark the two tcBF
    chords; the separate blue and orange segments mark the acBF chords.}
    \label{fig:aperture_chords}
\end{figure}

The angular integrals in Eqs.~\eqref{eq:tcbf_3d} and
\eqref{eq:acbf_3d} can be evaluated explicitly for a uniformly illuminated,
hard-edged circular aperture of semi-angle $\alpha$, with defocus as the
only aberration. The expressions below give the full complex transfer at
each $(q_x,q_y,q_z)$, not only its support boundaries.
Write
\begin{equation}
q_\perp=\sqrt{q_x^2+q_y^2}>0,
\qquad
\phi=\pi\lambda\Delta f\,q_\perp^2,
\qquad
\Omega_0=\pi\alpha^2.
\end{equation}
Resolve the detector angle into components $t\parallel\boldsymbol q$
and $u\perp\boldsymbol q$, so that
$\boldsymbol\Theta\cdot\boldsymbol q=q_\perp t$.
For either axial mapping, the delta function fixes $t$ and contributes
the Jacobian $1/q_\perp$:
\begin{equation}
\delta(q_z-q_\perp t+c)
=\frac{1}{q_\perp}
\delta\!\left(t-\frac{q_z+c}{q_\perp}\right).
\end{equation}
Here $c=0$ for tcBF and $c=\pm\lambda q_\perp^2/2$ for the
two acBF branches. The remaining integration over $u$ is geomtrically the length
of an aperture-overlap chord as illustrated in Figure~\ref{fig:aperture_chords}a.

Define the half-chord of one aperture by
\begin{equation}
h(t)=\sqrt{\max(\alpha^2-t^2,0)}.
\end{equation}
At fixed $q_\perp$, the double overlap (DO) and triple overlap (TO) chord lengths are
\begin{equation}
\begin{aligned}
\ell_\pm(t)&=2\min\{h(t),h(t\pm\lambda q_\perp)\},\\
\ell_0(t)&=2\min\{h(t),h(t-\lambda q_\perp),h(t+\lambda q_\perp)\}.
\end{aligned}
\label{eq:appendix_aperture_chord_lengths}
\end{equation}
Thus $\ell_0$ is the DSB chord, while $\ell_- -\ell_0$ and
$\ell_+ -\ell_0$ are the two SSB chord lengths.
The dependence of these lengths on $q_\perp$ is implicit.
Figure~\ref{fig:aperture_chords} illustrates the chord geometry and the
positions selected by the two axial mappings.

For tcBF, set $t=q_z/q_\perp$ in Eq.~\eqref{eq:tcbf_3d} to obtain
\begin{equation}
\PCTF_{\mathrm{tcBF}}^{\mathrm{3D}}(\boldsymbol q,q_z)
=\frac{i}{2\Omega_0q_\perp}
\left[\ell_-(t)e^{-i\phi}-\ell_+(t)e^{i\phi}\right].
\label{eq:appendix_tcbf_3d_closed_form}
\end{equation}
This includes both the signed real transfer and the imaginary
sideband imbalance. No angular integration remains.

For acBF, the two Ewald mappings select different chord positions,
\begin{equation}
t_\pm=\frac{q_z\mp\lambda q_\perp^2/2}{q_\perp}.
\label{eq:acbf_chord_positions}
\end{equation}
In an SSB region, the corrected branch weight is $1/(2\Omega_0)$.
In the DSB region, the combined per-tilt PCTF is proportional to
$\sin\phi$, so the common correction factor is
$\operatorname{sgn}(\sin\phi)$. Consequently,
\begin{equation}
\begin{aligned}
\PCTF_{\mathrm{acBF}}^{\mathrm{3D}}(\boldsymbol q,q_z)
=\frac{1}{2\Omega_0q_\perp}\Big\{&
\ell_-(t_-)-\ell_0(t_-)
+\ell_+(t_+)-\ell_0(t_+)\\
&+i\operatorname{sgn}(\sin\phi)
\left[\ell_0(t_-)e^{-i\phi}-\ell_0(t_+)e^{i\phi}\right]
\Big\}.
\end{aligned}
\label{eq:appendix_acbf_3d_closed_form}
\end{equation}
At an exact DSB transfer zero, the phase correction is undefined;
we suppress that contribution, corresponding to
$\operatorname{sgn}(0)=0$ in Eq.~\eqref{eq:appendix_acbf_3d_closed_form}.
At $q_\perp=0$, the original first-order phase-transfer expression
vanishes by cancellation of the two sidebands; the formulas containing
$1/q_\perp$ should not be evaluated there.

Equations~\eqref{eq:appendix_tcbf_3d_closed_form} and
\eqref{eq:appendix_acbf_3d_closed_form} reproduce
Eqs.~\eqref{eq:tcbf_3d_closed_form} and \eqref{eq:acbf_3d_closed_form}
in the main text and evaluate the stated angular-integral models
directly from $(q_x,q_y,q_z)$ and the optical parameters.
Their axial integrals recover the 2D projections, Eqs.~\eqref{eq:tcbf_2d_collapse} and
\eqref{eq:acbf_2d_collapse}, respectively.
The projection check acts on the complex transfer, not its magnitude;
it is a check of the 3D expressions rather than a derivation of their
axial support from the 2D limit.
The transfer maps in Figs.~\ref{fig:single_atom_analytic_ctf} and
\ref{fig:tcbf_acbf_3d_ctf_boundaries} were evaluated by numerical
detector-angle summation, rather than by these closed forms.
For general aberrations or a nonuniform aperture amplitude, the weights
can vary along each chord and a residual numerical integral is generally
required.

\subsection{Real, imaginary, and magnitude representations}

The three-dimensional PCTFs are generally complex because the two sidebands
are separated in $q_z$. Their imaginary components need not cancel at a fixed
$q_z$, even though the projected two-dimensional PCTFs are real.
For the tcBF representation in Eq.~\eqref{eq:tcbf_3d},
\begin{equation}
\PCTF_{\mathrm{tcBF}}^{\mathrm{3D}}(\boldsymbol q,q_z)
=
iL_-(\boldsymbol q,q_z)e^{-i\phi}
-
iL_+(\boldsymbol q,q_z)e^{+i\phi},
\end{equation}
where the angle-integrated, non-negative branch weights are
\begin{equation}
L_\pm(\boldsymbol q,q_z)
=
\frac{1}{2\Omega_0}\int \dd^2\boldsymbol\Theta\,
A(\boldsymbol\Theta)A(\boldsymbol\Theta\pm\lambda\boldsymbol q)
\delta(q_z-\boldsymbol\Theta\cdot\boldsymbol q)
=\frac{\ell_\pm(q_z/q_\perp)}{2\Omega_0q_\perp}.
\end{equation}
With $\phi=\pi\lambda\Delta f|\boldsymbol q|^2$ as above, the real and
imaginary parts describe signed in-phase transfer and sideband imbalance:
\begin{equation}
\begin{aligned}
\mathfrak{Re}\PCTF_{\mathrm{tcBF}}^{\mathrm{3D}}&=(L_-+L_+)\sin\phi,\\
\mathfrak{Im}\PCTF_{\mathrm{tcBF}}^{\mathrm{3D}}&=(L_--L_+)\cos\phi.
\end{aligned}
\end{equation}
For a centro-symmetric aperture, the change of variables
$\boldsymbol\Theta\rightarrow-\boldsymbol\Theta$ interchanges the aperture
products and reverses $\boldsymbol\Theta\cdot\boldsymbol q$.
Thus the angle-integrated branch weights obey
$L_-(\boldsymbol q,q_z)=L_+(\boldsymbol q,-q_z)$, so their
difference is antisymmetric in $q_z$.
Consequently, $\int\mathfrak{Im}\PCTF_{\mathrm{tcBF}}^{\mathrm{3D}}\dd q_z=0$:
the imaginary part cancels in the projected 2D PCTF, but can remain locally
where the branches are imbalanced, contributing to an asymmetric axial
point response.

For acBF, the SSB weights and the common DSB phase correction must both
be retained. Expanding Eq.~\eqref{eq:appendix_acbf_3d_closed_form} gives
\begin{equation}
\begin{aligned}
\mathfrak{Re}\PCTF_{\mathrm{acBF}}^{\mathrm{3D}}
={}&\frac{1}{2\Omega_0q_\perp}\Big\{
\ell_-(t_-)+\ell_+(t_+)-\ell_0(t_-)-\ell_0(t_+)\\
&\qquad{}+|\sin\phi|[\ell_0(t_-)+\ell_0(t_+)]\Big\},\\
\mathfrak{Im}\PCTF_{\mathrm{acBF}}^{\mathrm{3D}}
={}&\frac{\operatorname{sgn}(\sin\phi)\cos\phi}{2\Omega_0q_\perp}
[\ell_0(t_-)-\ell_0(t_+)].
\end{aligned}
\label{eq:appendix_acbf_real_imag}
\end{equation}
The imaginary part is confined to the DSB contribution and is odd in
$q_z$ for the circular aperture; its axial integral also vanishes.
These expressions accompany the complex-transfer plots in
Fig.~\ref{fig:tcbf_acbf_3d_ctf_boundaries}. The magnitude in
Fig.~\ref{fig:single_atom_analytic_ctf} discards the phase relationships
and must not replace the complex transfer in the $q_z$ projection check.


\subsection{Interpretation of the 3D PCTFs}

The 3D PCTF describes which axial spatial frequencies $q_z$ are transferred
for each transverse frequency $\boldsymbol q$. It should not be
interpreted only as a brightness map: the support geometry determines the
axial bandwidth, whereas the complex phase determines the signed transfer and
axial asymmetry.

For tcBF, the parallax mapping $q_z=\boldsymbol\Theta\cdot\boldsymbol q$
makes the axial bandwidth grow approximately linearly with
$|\boldsymbol q|$. At high transverse frequency, tcBF can transfer
relatively high $|q_z|$ components.

For acBF, the support remains on the Ewald sheets in
Eq.~\eqref{eq:ewald_sheets}. For equal circular apertures, its envelope in the
$q_y=0$ slice narrows as $|q_x|$ approaches $2\alpha/\lambda$. Thus acBF has less
high-$q_z$ support at high transverse frequency than tcBF. The tradeoff is
between different transfer properties: tcBF provides broader high-$q_z$
parallax support but retains stronger signed oscillations and axial phase
asymmetry, whereas acBF gives a phase-aligned projected transfer that remains
nonzero up to $2\alpha/\lambda$ but has a narrower high-$q_z$ envelope.

%% file: appendix_support_boundaries.tex
\FloatBarrier
\renewcommand{\thefigure}{C\arabic{figure}}
\renewcommand{\theHfigure}{C.\arabic{figure}}
\setcounter{figure}{0}
\renewcommand{\theequation}{C\arabic{equation}}
\renewcommand{\theHequation}{C.\arabic{equation}}
\setcounter{equation}{0}
\renewcommand{\thetable}{C\arabic{table}}
\renewcommand{\theHtable}{C.\arabic{table}}
\setcounter{table}{0}
\section[Analytical boundaries from the Ewald term]{Analytical boundaries in the plotted \texorpdfstring{$q_y=0$}{q-y=0} slice}
\label{appendix_B}

For the boundary derivation, we use dimensionless coordinates
\begin{equation}
\bar{\boldsymbol q}=\frac{\lambda\boldsymbol q}{\alpha},
\qquad
\bar q_z=\frac{\lambda q_z}{\alpha^2},
\qquad
\bar{\boldsymbol\Theta}=\frac{\boldsymbol\Theta}{\alpha}.
\end{equation}
In the $q_y=0$ section, write $\bar q=\bar q_x$.
The circular aperture becomes the unit disk $|\bar{\boldsymbol\Theta}|<1$.
By inversion symmetry it is sufficient to
derive the branch intervals for $0<\bar q<2$; the negative-frequency side
follows by inversion.

\subsection{tcBF boundaries}

For tcBF, $\bar q_z=\bar\Theta_x\bar q$.
The overlap $A(\alpha\bar{\boldsymbol\Theta})
A[\alpha(\bar{\boldsymbol\Theta}\pm\bar{\boldsymbol q})]\ne0$
requires
\begin{equation}
\max(-1,\mp\bar q-1)<\bar\Theta_x<\min(1,\mp\bar q+1).
\end{equation}
The signs here and below are paired with the corresponding aperture branch.
For $0<\bar q<2$, multiplying these limits by $\bar q$ gives
\begin{equation}
\begin{aligned}
-\bar q+\bar q^2&<\bar q_z<\bar q
&&\text{(minus branch)},\\
-\bar q&<\bar q_z<\bar q-\bar q^2
&&\text{(plus branch)}.
\end{aligned}
\end{equation}
These intervals define the parallax wedge and its quadratic branch-overlap
boundaries.

\subsection{acBF full-support boundaries}

For acBF, the branch mappings are
\begin{equation}
\bar q_z^\pm=\bar\Theta_x\bar q\pm\frac{\bar q^2}{2}.
\end{equation}
The full acBF real-part support is obtained from the same two-disk overlaps as tcBF, but mapped onto these Ewald sheets.

For equal circular apertures, the projected outer envelopes of the plus and minus sheets coincide. For $|\bar q|<2$,
\begin{equation}
{
\bar q_{z}^{\mathrm{full,upper/lower}}
=
\pm\left(|\bar q|-\frac{\bar q^2}{2}\right).
}
\end{equation}
This full support closes at $|\bar q|=2$.

\subsection{acBF DSB boundaries}

The acBF imaginary part is dominated by the double-sideband region. The DSB region requires triple overlap:
\begin{equation}
A(\alpha\bar{\boldsymbol\Theta})
A[\alpha(\bar{\boldsymbol\Theta}-\bar{\boldsymbol q})]
A[\alpha(\bar{\boldsymbol\Theta}+\bar{\boldsymbol q})]
\neq0.
\end{equation}
For $\bar q_y=0$, this gives
\begin{equation}
|\bar\Theta_x|<1-|\bar q|,
\qquad |\bar q|<1.
\end{equation}
Mapping this interval onto the two Ewald sheets gives all four DSB
boundaries in one expression:
\begin{equation}
-(|\bar q|-\bar q^2)
<\bar q_z^\pm\mp\frac{\bar q^2}{2}
<|\bar q|-\bar q^2,
\qquad |\bar q|<1.
\end{equation}
These four DSB curves set the main boundary of the acBF imaginary component.

Thus tcBF has broader high-$\bar q_z$ support at high transverse frequency because
its axial mapping follows parallax. acBF preserves the Ewald-sheet geometry
and phase-aligns the full per-tilt PCTF, producing smoother projected phase
transfer but a narrower high-$\bar q_z$ bandwidth at large $|\bar q_x|$.

%% file: appendix_confocal_comparison.tex
\renewcommand{\thefigure}{D\arabic{figure}}
\renewcommand{\theHfigure}{D.\arabic{figure}}
\setcounter{figure}{0}
\renewcommand{\theequation}{D\arabic{equation}}
\renewcommand{\theHequation}{D.\arabic{equation}}
\setcounter{equation}{0}
\renewcommand{\thetable}{D\arabic{table}}
\renewcommand{\theHtable}{D.\arabic{table}}
\setcounter{table}{0}

\section{3D information beyond the bright field}
\label{appendix_C_confocal_comparison}

Xin and Muller derived 3D CTFs for focal-series STEM and SCEM before the
development of parallax ptychographic depth reconstruction
\citep{xin_2010_ThreeDimensionalImagingAberrationCorrected}. The key insight of
Ref.~\citep{terzoudis-lumsden_2023_ResolutionVirtualDepth} was an approximate
mapping from such a ptychographic reconstruction to an ideal confocal system
with a hypothetical phase-sensitive point detector. When the postspecimen
collection semi-angle $\beta$ is larger than the probe-forming semi-angle
$\alpha$, the mapped 3D pseudo-CTF contains additional butterfly-shaped
support.

This ideal confocal result is a useful bound on the support that could be
available after complex-field recovery, but it is not a quantitative CTF for
ptychographic data. The mapping assumes uniform illumination of the lower
postspecimen aperture, whereas that aperture is populated and recovered with
nonuniform weights in ptychography. It also takes the phase of the recovered
complex field as its observable. The larger support is therefore not obtained
by retaining a more complete set of first-order WPOA intensity terms. Relative
to the recorded 4D-STEM intensity, the construction combines the weak-phase
object expansion with a complex-field recovery step outside the first-order
intensity expansion. tcBF and acBF instead act linearly on the measured
detector intensities.

\subsection{Same object approximation, different observables}

For a projected potential $V(\mathbf r)$ and electron--specimen interaction
constant $\sigma$,
\begin{equation}
e^{i\sigma V(\mathbf r)}
=
1+i\sigma V(\mathbf r)+O(V^2).
\label{eq:appendix_c_wpoa}
\end{equation}
We use the position--angle notation of the main text:
$\boldsymbol\Theta$ is detector angle,
$\boldsymbol q$ is transverse object spatial frequency in cycles per unit
length, and $\lambda\boldsymbol q$ is its angular separation.
The two aperture functions are
\begin{equation}
A_\alpha(\boldsymbol\Theta)
=\operatorname{circ}\!\left(\frac{|\boldsymbol\Theta|}{\alpha}\right),
\qquad
A_\beta(\boldsymbol\Theta)
=\operatorname{circ}\!\left(\frac{|\boldsymbol\Theta|}{\beta}\right).
\label{eq:appendix_c_apertures}
\end{equation}

In the ideal mapped confocal calculation, the complex scattering amplitudes
across the lower aperture are treated as uniformly available before the phase
is taken. Its linear pseudo-CTF therefore contains aperture products of the
form
\begin{equation}
A_\alpha(\boldsymbol\Theta)
A_\beta(\boldsymbol\Theta\mp\lambda\boldsymbol q).
\label{eq:appendix_c_confocal_apertures}
\end{equation}
One member of the interfering pair is limited by $\alpha$, while the other can
extend to $\beta$. The lateral support consequently reaches
$|\boldsymbol q|<(\alpha+\beta)/\lambda$. For $\beta>\alpha$, the interval beyond
$2\alpha/\lambda$ forms the four additional wings in the three-dimensional
pseudo-CTF. This contribution is linear in $V$ because the relative phase of
the two complex amplitudes is part of the reconstructed observable. This
linearity describes the phase of the recovered complex field; it does not
establish a linear dark-field term in the recorded intensity.

The recorded 4D-STEM data are instead intensities. At one detector pixel,
the wave and measured intensity expand as
\begin{equation}
\begin{aligned}
\Psi
&=
\Psi^{(0)}+\Psi^{(1)}+O(V^2),
\qquad
\Psi^{(1)}=O(V),
\\
|\Psi|^2
&=
|\Psi^{(0)}|^2
+2\mathfrak{Re}\!\left[\Psi^{(0)*}\Psi^{(1)}\right]
+O(V^2).
\end{aligned}
\label{eq:appendix_c_intensity_expansion}
\end{equation}
The first term is the reference intensity
$I_0=|\Psi^{(0)}|^2$ used in Eq.~\eqref{eq:detector_resolved_contrast}; the
second is the linear, object-dependent phase contrast retained in $D$. This
interference term requires the unscattered amplitude $\Psi^{(0)}$ as a phase
reference and therefore exists only inside the bright-field disk.

For a collection angle $\beta\ge\alpha$, each first-order detector sideband
contains
\begin{equation}
\begin{aligned}
A_\beta(\boldsymbol\Theta)
A_\alpha(\boldsymbol\Theta)
A_\alpha(\boldsymbol\Theta\mp\lambda\boldsymbol q)
&=
A_\alpha(\boldsymbol\Theta)
A_\alpha(\boldsymbol\Theta\mp\lambda\boldsymbol q).
\end{aligned}
\label{eq:appendix_c_collection_cancels}
\end{equation}
The collection aperture therefore drops out once it contains the bright-field
disk. The two remaining probe-aperture factors restrict the linear lateral
support to $|\boldsymbol q|<2\alpha/\lambda$. Linear shift correction, phase
correction, weighting, and summation of the detector-pixel contrasts do not
alter this aperture product.

Outside the bright-field disk, $A_\alpha(\boldsymbol\Theta)=0$. The resulting
diffraction-intensity ordering is
\begin{equation}
2\mathfrak{Re}\!\left[\Psi^{(0)*}\Psi^{(1)}\right]=0,
\qquad
|\Psi^{(1)}|^2+O(V^3)
\sim V^2.
\label{eq:appendix_c_df_order}
\end{equation}
Dark-field intensity may be useful, but it is not part of a consistently
first-order intensity CTF. Iterative ptychography can nevertheless exploit the
nonlinear dark-field signal to recover additional lateral and axial spatial
frequencies, and thereby achieve higher resolution in both directions.
Because directly measured dark-field intensity has no linear phase CTF,
recovering this phase-resolved three-dimensional information generally
requires an iterative reconstruction algorithm
\citep{ma_2026_Information4DSTEMWhere}.

Thus, when the unequal-pupil pseudo-CTF is interpreted as transfer from the
recorded intensity, the weak-phase object expansion has been combined with a
complex-field phase-recovery step whose nonlinear intensity dependence is not
included in that first-order expansion. The butterfly is therefore not a
more complete first-order WPOA intensity result.

\subsection{Comparison of the three-dimensional transfer support}

The two 3D CTFs differ because they describe different observables and retain
different orders of the measured intensity. The reconstructed phase in
Ref.~\citep{terzoudis-lumsden_2023_ResolutionVirtualDepth} is an $O(V)$
quantity formed after complex-field recovery and retains the unequal-pupil
product in Eq.~\eqref{eq:appendix_c_confocal_apertures}. The measured tcBF/acBF
phase contrast is the $O(V)$ interference term in
Eq.~\eqref{eq:appendix_c_intensity_expansion} and is limited by the equal-pupil
product in Eq.~\eqref{eq:appendix_c_collection_cancels}. Measured elastic
dark-field intensity begins at $O(V^2)$.

\begin{figure}[!htbp]
    \centering
    \includegraphics[width=\linewidth]{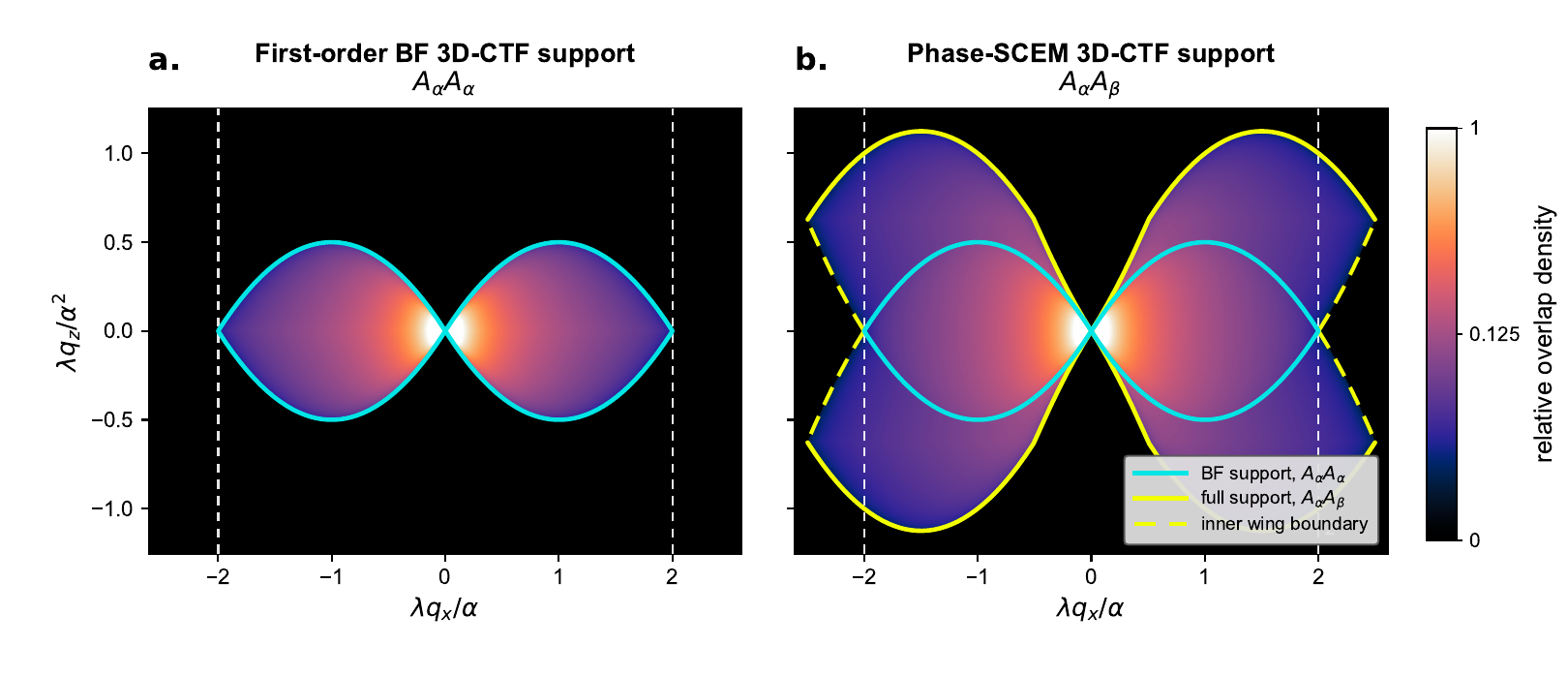}
    \caption{Comparison of 3D contrast-transfer support in the $q_y=0$ plane
for 300~keV electrons, with illumination and collection semi-angles
$\alpha=20$~mrad and $\beta=30$~mrad, respectively.
The axes are normalized as $\lambda q_x/\alpha$ and
$\lambda q_z/\alpha^2$, as in Fig.~\ref{fig:tcbf_acbf_3d_ctf_boundaries}.
    (a) Equal-pupil overlap $A_\alpha A_\alpha$ for the first-order
    interference term measured in bright-field.
    (b) Unequal-pupil overlap $A_\alpha A_\beta$ in the ideal phase-sensitive
    confocal mapping of Ref.~\citep{terzoudis-lumsden_2023_ResolutionVirtualDepth}.
    Its support reaches $|\lambda q_x/\alpha|=1+\beta/\alpha=2.5$, forming four wings
    beyond the bright-field support. The cyan contour marks the bright-field
    boundary in both panels; solid and dashed yellow contours mark the outer
    and inner boundaries of the additional support in (b). White dashed lines
    mark $|\lambda q_x/\alpha|=2$. Color shows nonnegative aperture-overlap
    density on a common, cube-root display scale. The extra support bounds
    what may be available after full complex-field recovery; it is not a
    linear dark-field term in the measured intensity or a quantitative
    ptychographic CTF.}
    \label{fig:butterfly_comparison}
\end{figure}

Figure~\ref{fig:butterfly_comparison} compares the aperture-overlap
support in the $q_y=0$ section of the 3D transfer function for 300~keV
electrons with $\alpha=20$~mrad and $\beta=30$~mrad.
The coordinates $q_x,q_y,q_z$ are spatial frequencies in cycles per unit
length, consistent with the main text and \ref{appendix_A_rose_pctf}.
%

\FloatBarrier

\begin{table}[htbp]
\centering
\small
\renewcommand{\arraystretch}{1.18}
\caption{Observable and leading object-dependent order.}
\label{tab:appendix_c_order_comparison}
\begin{tabularx}{\textwidth}{>{\raggedright\arraybackslash}X|>{\centering\arraybackslash}X|>{\raggedright\arraybackslash}X}
\hline
Quantity & Leading order & Effect of $\beta>\alpha$ \\
\hline
Reconstructed complex-field phase
& $O(V)$
& Adds unequal-pupil support and the four wings \\
\hline
Measured BF interference
$2\mathfrak{Re}[\Psi^{(0)*}\Psi^{(1)}]$
& $O(V)$
& No extension after $\beta\ge\alpha$ \\
\hline
Measured elastic DF intensity $|\Psi^{(1)}|^2$
& $O(V^2)$
& Adds nonlinear contrast transfer outside the BF disk \\
\hline
\end{tabularx}
\end{table}

Ref.~\citep{terzoudis-lumsden_2023_ResolutionVirtualDepth} therefore provides
an approximate mapping from parallax ptychography to the phase CTF of an ideal
confocal field. The resulting butterfly is a useful upper bound, but the
uniform lower-pupil illumination and ideal complex-field observable prevent it
from being a quantitative ptychographic intensity CTF. It does not arise from
a more complete first-order detector-intensity expansion. In a consistent
first-order expansion of the directly measured 4D-STEM intensity, the
dark-field term vanishes at $O(V)$ and begins at $O(V^2)$. This limits the
linear phase CTF, not the total information in the dark-field data: iterative
ptychography can recover additional lateral and axial information from the
nonlinear signal.

%% file: appendix_spherical_aberration.tex
\FloatBarrier
\renewcommand{\thefigure}{E\arabic{figure}}
\renewcommand{\theHfigure}{E.\arabic{figure}}
\setcounter{figure}{0}
\renewcommand{\theequation}{E\arabic{equation}}
\renewcommand{\theHequation}{E.\arabic{equation}}
\setcounter{equation}{0}

\section{3D CTF from higher-order aberrations}
\label{appendix_E_aberrations}

We extend the analytical calculations in
Figs.~\ref{fig:single_atom_analytic_ctf} and
\ref{fig:tcbf_acbf_3d_ctf_boundaries} to a probe with
spherical aberration. With the same phase convention as the main text,
\begin{equation}
\chi(\boldsymbol\Theta)
=-\frac{k_0\Delta f}{2}|\boldsymbol\Theta|^2
+\frac{k_0 C_s}{4}|\boldsymbol\Theta|^4,
\qquad
\Delta\mathbf r_\chi
=\bigl(-\Delta f+C_s|\boldsymbol\Theta|^2\bigr)\boldsymbol\Theta.
\label{eq:cs_phase_shift}
\end{equation}
Here $C_s$ is the spherical-aberration coefficient. tcBF registration uses
both terms in the shift, following Eq.~\eqref{eq:general_aberration_shift}.

The terms beyond the common shift can be seen by expanding
$\chi(\boldsymbol\Theta\pm\lambda\boldsymbol q)$ about each detector angle.
The local quadratic term is set by
\begin{equation}
\frac{1}{k_0}\nabla_{\boldsymbol\Theta}\nabla_{\boldsymbol\Theta}\chi
=\bigl(-\Delta f+C_s|\boldsymbol\Theta|^2\bigr)\mathbb I
+2C_s\boldsymbol\Theta\boldsymbol\Theta^{\mathsf T},
\label{eq:cs_phase_curvature}
\end{equation}
where $\mathbb I$ is the two-dimensional identity matrix.
Its tangential and radial eigenvalues are
$-\Delta f+C_s|\boldsymbol\Theta|^2$ and
$-\Delta f+3C_s|\boldsymbol\Theta|^2$, respectively.
These give a local defocus and astigmatism that vary with angle; higher
powers of $\boldsymbol q$ remain at finite spatial frequency.
A shift alone cannot correct this contrast. acBF uses the full
per-angle PCTF phase, including both overlapping sidebands
[Eq.~\eqref{eq:acbf_correction_phasor}].

\begin{figure}[!htbp]
\centering
\includegraphics[width=.95\textwidth]{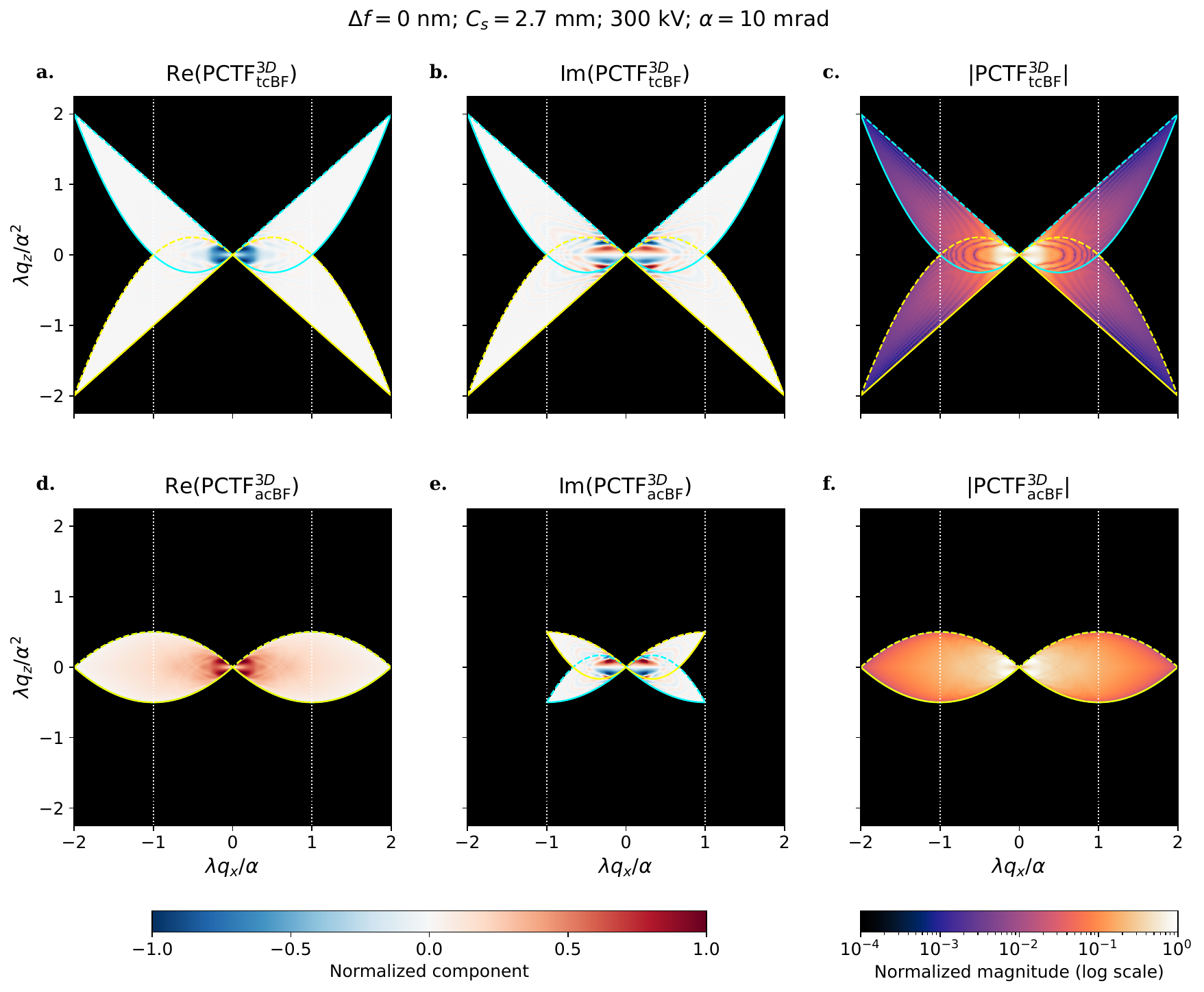}
\caption{Complex analytical 3D PCTFs at zero defocus with $C_s=2.7$~mm,
300~kV, and $\alpha=10$~mrad. The top row (a--c) shows tcBF and the
bottom row (d--f) shows acBF; columns show the real component, imaginary
component, and magnitude $|\mathrm{PCTF}^{3D}|$, respectively.
Each panel is normalized separately to its maximum absolute value within
its support.
Cyan and yellow curves
show the sideband envelopes, which coincide for the real part and magnitude of acBF. Thec imaginary part of acBF is instead restricted to
the double-sideband region ($<1\alpha$.)}
\label{fig:cs_complex_zero}
\end{figure}

\begin{figure}[!htbp]
\centering
\includegraphics[width=.95\textwidth]{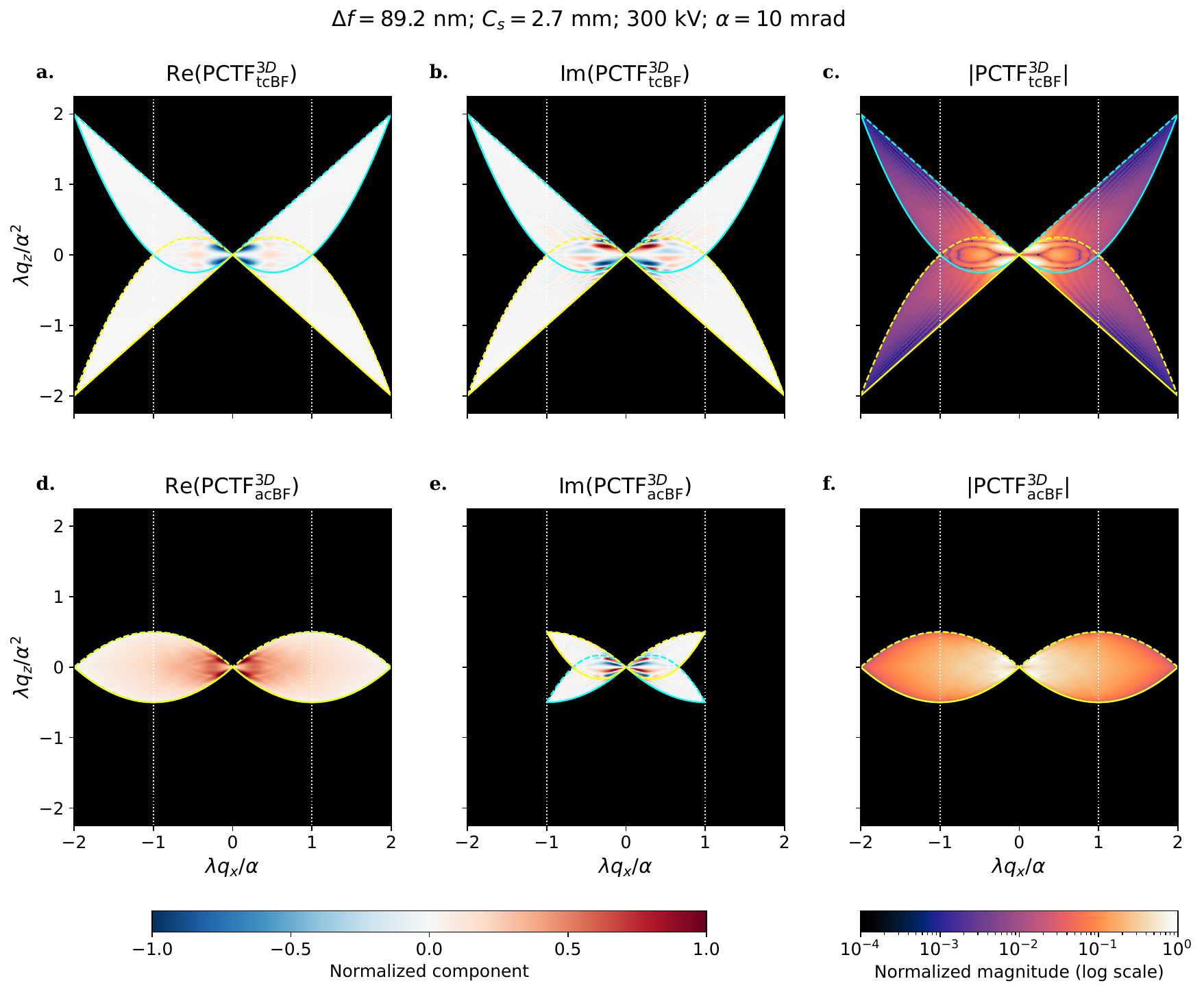}
\caption{Complex analytical 3D PCTFs at $\Delta f=89.2$~nm, with the same
 $C_s=2.7$~mm,
300~kV, and $\alpha=10$~mrad as in Fig.~\ref{fig:cs_complex_zero}.
Changing defocus alters the transfer inside the envelopes, while their
geometric boundaries remain unchanged.}
\label{fig:cs_complex_scherzer}
\end{figure}

Figures~\ref{fig:cs_complex_zero} and \ref{fig:cs_complex_scherzer} show
the real and imaginary components and magnitudes for 300~kV, $\alpha=10$~mrad, and
$C_s=2.7$~mm, with $\Delta f=0$ and 89.2~nm (near Scherzer in our sign
convention). $C_s$ is held fixed while depth varies through defocus.
It changes the complex weights, not the aperture overlaps or depth-frequency
mappings of \ref{appendix_B}. The geometric envelopes therefore match the
defocus-only case at the same $\alpha$, without implying uniform or nonzero
transfer inside them.

Residual phases can cancel in tcBF after registration. acBF phase alignment
reduces cancellation but retains the magnitudes and zeros within the TO region, so its axial
response can still depend on $C_s$. These separately normalized, weak-phase
calculations do not establish a resolution gain; that requires a matched
axial-response or edge test, as discussed in
Sec.~\ref{sec:higher_order_depth}.

\FloatBarrier

%% file: appendix_aberration_sinograms.tex
\FloatBarrier
\renewcommand{\thefigure}{F\arabic{figure}}
\renewcommand{\theHfigure}{F.\arabic{figure}}
\setcounter{figure}{0}
\renewcommand{\theequation}{F\arabic{equation}}
\renewcommand{\theHequation}{F.\arabic{equation}}
\setcounter{equation}{0}

\section{Parallax from higher-order aberrations of different parity}
\label{appendix_F_sinograms}

\begin{figure}[!p]
\centering
\includegraphics[width=\textwidth,height=.9\textheight,keepaspectratio]{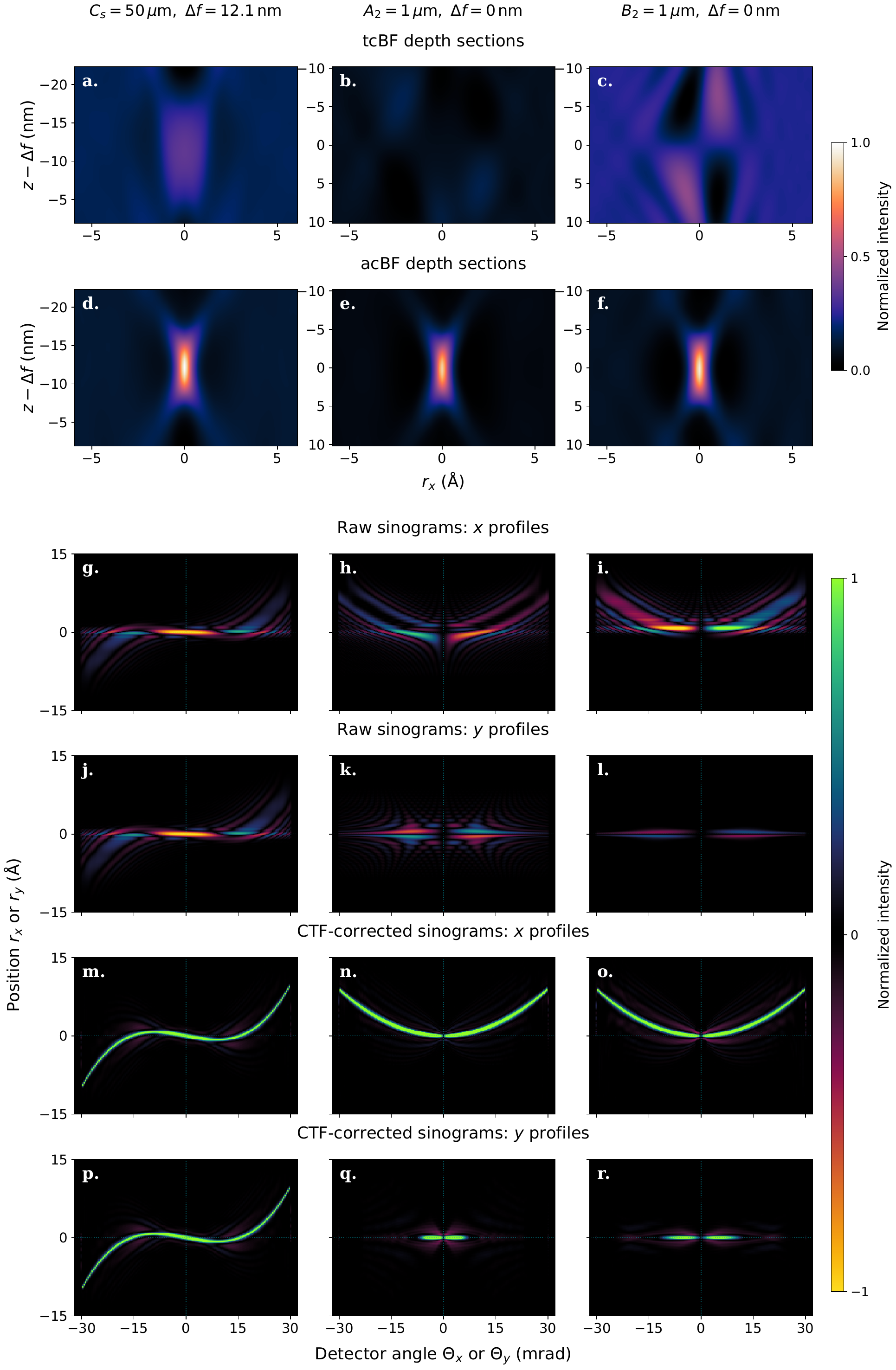}
\caption{Single C atom depth sections and sinograms with $C_s$, $A_2$,
and $B_2$ (columns).
(a--c) tcBF and (d--f) acBF depth sections in the $r_x$--$\Delta z$ plane,
where $\Delta z=z-\Delta f$.
(g--i,j--l) Raw central $x$ and $y$ sinograms.
(m--o,p--r) CTF-corrected sinograms with geometrical
parallax retained.
Even aberration phases give antisymmetric shift trajectories, whereas
odd phases give symmetric trajectories.
}
\label{fig:aberration_sinograms}
\end{figure}

Figure~\ref{fig:aberration_sinograms} compares single C atom simulations
with spherical aberration $C_s$, threefold astigmatism $A_2$, and axial
coma $B_2$.
Simulations used 300~kV, $\alpha=30$~mrad,
a $200\times200$ scan at 0.15~\AA{} sampling,
and $166\times166$ detector pixels at 0.395~mrad sampling.
The coefficients are $C_s=5\times10^5$~\AA{} with $\Delta f=121$~\AA{},
or $A_2=B_2=10^4$~\AA{} separately at zero defocus.
The first two rows show tcBF and acBF depth sections reconstructed by sweeping $z$ over a 200~\AA{} interval in
4~\AA{} increments. The plotted axial coordinate is 
$\Delta z=z-\Delta f$.
Both use the full aberration-phase gradient for the virtual-image shifts,
including the specified higher-order aberration, rather than defocus alone.
tcBF registers and sums the virtual images; acBF additionally corrects
the residual phase contrast transfer.
Sinograms are central line cuts through the 4D dataset and directly reveal
the parallax information (Fig.~\ref{fig:single_atom_sinograms}).

\subsection{Phase parity determines shift parity}

The relevant symmetry is inversion of detector angle,
$\boldsymbol\Theta\rightarrow-\boldsymbol\Theta$.
Using the shift defined in Eq.~\eqref{eq:general_aberration_shift},
$\Delta\mathbf r_\chi=k_0^{-1}\nabla_{\boldsymbol\Theta}\chi$, gives
\begin{equation}
\chi(-\boldsymbol\Theta)=\pm\chi(\boldsymbol\Theta)
\quad\Longrightarrow\quad
\Delta\mathbf r_\chi(-\boldsymbol\Theta)
=\mp\Delta\mathbf r_\chi(\boldsymbol\Theta).
\label{eq:aberration_shift_parity}
\end{equation}
The upper sign denotes an even phase and the lower sign an odd phase.
Thus an even phase produces opposite shifts at opposite detector
angles, whereas an odd phase produces the same shift.
Differentiation reverses parity; the two odd aberrations need not share
the same rotational symmetry.

Defocus and spherical aberration have even phases.
Their shift is given by Eq.~\eqref{eq:cs_phase_shift}, which when projected along $x$ in the detector plane reduces to
\begin{equation}
\Delta r_{\chi,x}(\Theta_x,0)
=-\Delta f\,\Theta_x+C_s\Theta_x^3.
\label{eq:cs_sinogram_trajectory}
\end{equation}
Both terms reverse sign with $\Theta_x$, producing the antisymmetric
trajectory in the first column.
Rotational symmetry gives the same trajectory for the orthogonal line.

For the aberration orientations used here, the odd phases are
\begin{equation}
\begin{aligned}
\chi_{A_2}(\boldsymbol\Theta)
&=\frac{k_0 A_2}{3}\left(\Theta_x^3-3\Theta_x\Theta_y^2\right),\\
\chi_{B_2}(\boldsymbol\Theta)
&=\frac{k_0 B_2}{3}\Theta_x\left(\Theta_x^2+\Theta_y^2\right).
\end{aligned}
\label{eq:odd_aberration_phases}
\end{equation}
Their gradients give
\begin{equation}
\begin{aligned}
\Delta\mathbf r_{\chi_{A_2}}
&=A_2\left(\Theta_x^2-\Theta_y^2,\,-2\Theta_x\Theta_y\right),\\
\Delta\mathbf r_{\chi_{B_2}}
&=\frac{B_2}{3}\left(3\Theta_x^2+\Theta_y^2,\,2\Theta_x\Theta_y\right).
\end{aligned}
\label{eq:odd_aberration_shifts}
\end{equation}
Along $\Theta_y=0$, the longitudinal shifts are
$A_2\Theta_x^2$ and $B_2\Theta_x^2$.
They therefore bend in the same direction for either sign of detector
angle and coincide when $A_2=B_2$.
Their different angular symmetries appear on the orthogonal detector line:
at $\Theta_x=0$, both have zero longitudinal $y$ shift, but the
transverse $x$ shift is $-A_2\Theta_y^2$ for threefold astigmatism
and $+B_2\Theta_y^2/3$ for coma.
The atom image consequently leaves the fixed central $y$ profile as
$|\Theta_y|$ increases, reducing its visible signal.
A flat or fading central-line signal does not imply zero two-dimensional
shift.

\subsection{CTF correction restores the parallax}

The corrected sinograms use the residual transfer introduced in
Eq.~\eqref{eq:acbf_correction_phasor}, without applying registration,
\begin{equation}
\begin{aligned}
\PCTF^{\mathrm{res}}(\boldsymbol q,\boldsymbol\Theta)
&=e^{i(2\pi\boldsymbol q)\cdot\Delta\mathbf r_\chi}
\PCTF(\boldsymbol q,\boldsymbol\Theta),\\
\tilde D_{\mathrm{corr}}(\boldsymbol q,\boldsymbol\Theta)
&=\tilde D(\boldsymbol q,\boldsymbol\Theta)
e^{-i\arg\PCTF^{\mathrm{res}}(\boldsymbol q,\boldsymbol\Theta)}.
\end{aligned}
\label{eq:sinogram_residual_ctf_correction}
\end{equation}
where CTF correction is applied to each 2D virtual
image within the BF mask.
Pixels outside the bright-field mask are unchanged.

The correction removes contrast reversals while retaining the shift
specified by the phase gradient, revealing the parallax effect embedded in the 4D data
 that we can use for 3D reconstruction.
The retained shift trajectories are prescribed by the gradient of the aberration function 
predicted by Eq.~\eqref{eq:general_aberration_shift}.


\FloatBarrier